\pdfoutput=1

\documentclass[12pt,letterpaper]{article}

\usepackage{mystyle} 

\title{On the Impossibility of Convergence of Mixed Strategies with Optimal No-Regret Learning}
\author{Vidya Muthukumar%
\thanks{Equal Contribution.}
\thanks{School of Electrical \& Computer Engineering and School of Industrial \& Systems Engineering, Georgia Institute of Technology. This work done while the author was at Department of Electrical Engineering and Computer Sciences and Simons Institute for the Theory of Computing, University of California, Berkeley, CA 94720. {\tt\small vmuthukumar8@gatech.edu}}
\and Soham Phade%
\footnotemark[1]
\thanks{Salesforce Research. This work done while the author was at Department of Electrical Engineering and Computer Sciences, University of California, Berkeley, CA 94720.
        {\tt\small soham\_phade@berkeley.edu, sahai@eecs.berkeley.edu}}%
\and Anant Sahai%
\thanks{Department of Electrical Engineering and Computer Sciences, University of California, Berkeley, CA 94720}
}
\date{}

\begin{document}
\maketitle

%\tableofcontents
%\listoffigures
%\listoftables

%Abstract
%!TEX root = ../main.tex
%
\begin{abstract}
%\cb
% We study the limiting behavior of the mixed strategies that result from a general class of optimal no-regret learning strategies in a repeated game setting where the stage game is any $2 \times 2$ competitive game (for which all the Nash equilibria of the game are completely mixed) that may be zero-sum or non-zero-sum.
% We consider optimal no-regret strategies that are mean-based (i.e. information set at each step is the empirical average of the opponent's realized play) and monotonic (either non-decreasing or non-increasing) in their argument.
% We show that for \textit{any} such choice of strategies, the limiting mixed strategies of the players cannot converge almost surely to any Nash equilibrium.
% This negative result is also shown to hold under a broad class of relaxations of these assumptions, which includes popular variants of Online-Mirror-Descent with optimism and/or adaptive step-sizes.
% Finally, we conjecture that the monotonicity assumption can be removed, and provide partial evidence for this conjecture. 
% Our results identify the inherent stochasticity in players' realizations as a critical factor underlying this divergence, and demonstrate a crucial difference in outcomes between using the opponent's mixtures and realizations to make strategy updates.
We study the limiting behavior of the mixed strategies that result 
from optimal no-regret learning strategies in a repeated game setting 
where the stage game is any $2 \times 2$ competitive game. 
We consider optimal no-regret algorithms that are mean-based and monotonic in their argument. 
We show that for any such algorithm, 
the limiting mixed strategies of the players cannot converge almost surely to any Nash equilibrium. 
This negative result is also shown to hold under a broad relaxation of these assumptions, 
including popular variants of Online-Mirror-Descent with optimism and/or adaptive step-sizes. 
Finally, we conjecture that the monotonicity assumption can be removed, and provide partial evidence for this conjecture. 
Our results identify the inherent stochasticity in players’ realizations as a critical factor underlying this divergence, and demonstrate a crucial difference in outcomes between using the opponent’s mixtures and realizations to make updates.


\end{abstract}

% Text of your paper here
%Intro and related work
%!TEX root = ../main.tex

\section{Introduction}
\label{sec: intro}

The mixed strategy Nash equilibrium (NE) is one of the oldest solution concepts central to game theory.
A finer understanding of how the NE arises as an outcome of learning behavior in a repeated game setting continues to be an active area of research.
Classical research in economics dating back to~\citet{brown1951iterative} and~\citet{robinson1951iterative} as well as recent work in computer science~\citep{freund1999adaptive} 
tells us that when both the players in a two-player zero-sum game use strategies based on no-regret learning dynamics~\citep{hannan1957approximation, littlestone1989weighted,kalai2005efficient},
then the time-average of their strategies will converge, almost surely, to a Nash equilibrium~\citep{freund1999adaptive}.
However, the convergence of the \textit{time-averaged} mixed actions to a NE does not necessarily imply that the \textit{day-to-day behavior} (i.e. the sequence of mixed strategies) of these players converges.
The convergence of day-to-day behavior (or its lack thereof) is a fundamentally important object of study in multi-agent systems with applications to economics, markets, evolutionary games, and  multi-robot control systems.
Often, the individual learning agents in these settings are trained using online algorithms that are based on no-regret learning strategies---while the classic line of literature~\citep{freund1999adaptive,hart2005adaptive,foster1997calibrated} shows that their average behavior will converge to an equilibrium concept, their day-to-day behavior remains poorly understood.
% In this paper, we are interested in the convergence properties of the mixed strategies (and not just their time averages) when the players make use of no-regret learning strategies.
% Further, we will establish these convergence properties (rather, their lack of) as a fundamental consequence of the no-regret requirement (and a few other technical conditions) without limiting our analysis to any specific no-regret learning algorithm.

In the asymptotic sense, the quantity of interest is the tuple of the limiting mixed strategies of both players, also referred to as the last-iterate (e.g. in~\citet{daskalakis2018last}).
\citet{bailey2018multiplicative} discovered the following surprising property of the last iterate:
When the players in a two-player zero-sum game compete against each other with the popular \emph{multiplicative weights algorithm} (which satisfies the no-regret property), then their resulting mixed strategies drift away from any interior NE --- in fact, they drift towards the boundary of the strategy space.
This intriguing result is derived in an environment where players can play what we term \textit{telepathic strategies}, i.e. player $1$ can observe the exact mixed strategies used by player $2$, and vice versa.
%\footnote{In more detail: the nature of telepathic strategies means that~\citet{bailey2018multiplicative} consider a deterministic dynamic system comprised of the pair of mixed actions evolving according to the multiplicative weights updates on the time-average of the opponents mixed actions.
% In contrast, we are interested in the stochastic dynamic system of the pair of mixed actions whose evolution depends on the past realizations of the mixed actions.}.
However, in the traditional repeated game setting, players can only observe the \emph{realizations} of the opponent's mixed strategies.
One would only expect the oscillation problem to be exacerbated by the ensuing stochastic feedback.

The natural question that arises is whether these \textit{last-iterate oscillations} are a specific property of the family of multiplicative weights algorithms, or a fundamental consequence of the no-regret property itself.
This paper provides substantial evidence that it is the latter, by showing that last-iterate oscillation occurs for a broad, generic class of asymptotically optimal no-regret algorithms in the traditional repeated game setting (where players only observe realizations of each other's mixed strategies, not the mixed strategies themselves).
% We study the traditional repeated game setting in which players can only observe the realizations of the opponent's mixed strategies; thus, the strategies cannot be telepathic.
In this ``non-telepathic'' scenario, we show that the ensuing stochasticity in realizations is one of the critical ingredients underlying the last-iterate oscillation.
% {\color{blue} Our analysis is not limited to any specific no-regret algorithm or family of algorithms (e.g. the Online-Mirror-Descent family~\cite{nemirovsky1983problem}), but applies to a broad class of algorithms satisfying some purely qualitative conditions.}
Our analysis is not limited to explicit algorithms or family of algorithms, and suggests that last-iterate oscillation can arise as a fundamental consequence of the optimal-no-regret requirement (as well as a few other technical conditions, which we specify shortly).
In other words, \emph{the notions of optimal-no-regret and convergence of the limiting mixed strategies may inherently conflict with one another.}
% Our results suggest that learning algorithms possess certain intrinsic properties by which the two notions---optimal-no-regret and convergence of the limiting mixed strategies---inherently conflict with one another.

% \sph{Include a line explaining our approach, how it is different from using ODE analysis and the need for it because we do not focus on any particular no-regret algorithm.
% Instead we give a general fundamental result.
% }

{\bf Our contributions:} We consider a repeated $2 \times 2$ game, i.e. a two player game repeatedly played infinitely many times at steps $t = 1, 2, \dots$, where both the players can play mixtures of two pure strategies each.
The repeated game strategy for a player outlines the rule by which she picks her mixed action at step $t$ based on the history up to and including step $(t-1)$.
We will first describe our main result in its most stylized form to help identify the key components responsible for the phenomenon of last-iterate oscillation.
We will make three natural assumptions on each player's repeated game strategy, all of which are ubiquitous to explicitly defined learning dynamics in the literature:
\begin{enumerate}
\item We assume the player's strategy to be an \textit{optimal no-regret} strategy with respect to her utility function, that is, she has an expected average regret of $\mathcal{O}(t^{1/2})$ irrespective of the strategy employed by the other player.
See Definition~\ref{def: uniform_noregret_rate} for formal definitions of no-regret algorithms, optimal or otherwise.
\item We assume that the player's optimal no-regret strategy is \textit{mean-based}, i.e. the player uses only the empirical average of the actions of the other player at step $(t-1)$ as a sufficient statistic to decide her mixed action at step $t$.
In other words, the player is agnostic to the ordering in the opponent's action realizations. 
Note, further, that such strategies are \textit{self-agnostic}, in the sense that they do not use the actual realizations of their own mixed strategies to update their strategy.
In general, the player is aware of the step $t$, and we accordingly allow her rule for mapping empirical averages to mixed strategies to depend on the step $t$.
We also do not require this assumption in an exact sense---in Section~\ref{sec: beyond_mean_based}, we show the validity of our results with approximately-mean-based strategies that display recency bias.
\item We assume that the player's optimal no-regret, mean-based strategy is \textit{monotonic} in its argument, i.e. the empirical average of her opponent's actions, at every step $t \geq 1$.
The monotonicity does not need to be strict, and its direction (increasing or decreasing) can vary arbitrarily across rounds.
We note that even the special case of monotonicity in the direction of the player's best response on all rounds (in the sense that the larger the relative advantage of a response is, the more likely a player is to use it) is a natural constraint to impose on a rational agent.
In this context, time-varying monotonicity constitutes a significantly weaker regularity condition on the player's strategies.
We discuss a possible relaxation of this monotonicity assumption in Section~\ref{sec:conjecture}.
% \footnote{In fact, this flexibility is in a certain sense \textit{required} for the property of no-regret to hold;  in particular, it can be shown that no mean-based rule that is the same across all steps admits a no-regret strategy.}
% \item We assume that a player's strategy is \textit{self-agnostic}\footnote{A self-agnostic repeated game strategy has the following useful property: if it is a no-regret strategy with respect to an oblivious opponent, then it is a no-regret strategy with respect to a non-oblivious opponent.
% See Chapter $4$,~\citep{cesa2006prediction} for definitions of oblivious and non-oblivious opponents.}, i.e. it does not use the actual realizations of her own mixed actions to update her strategy.
% In other words, the player picks her mixed action at step $t$ only based on the action realizations of the other player up to, and including, step $(t-1)$.
\end{enumerate}

Most popular online learning dynamics, such as \textit{Online-Mirror-Descent}~\citep{nemirovsky1983problem} (or \textit{Follow-the-Regularized Leader}~\cite{shalev2011online}) strategies, can easily be verified to satisfy all three of these assumptions.
Our main contribution is to show that if both players deploy strategies satisfying the above three properties, and the stage game possesses a unique
completely mixed NE\footnote{We note that any $2\times2$ game that possesses only completely mixed NE, can be shown to possess one unique NE, which is also the unique correlated equilibrium of the game~\citep{phade2019geometry}.
These games are designated as \textit{competitive games} in an unpublished manuscript by~\citet{calvo2006set}. 
These games have been of special interest in the design of experiments to test the performance of NE as a predictor of behavior in games as they have an unambiguous NE. For example,~\citet{selten2008stationary,binmore2001does} refer to these games as completely mixed $2\times 2$ games.}, 
then their mixed actions \textit{cannot} converge to Nash NE.
% The natural question arises whether the resulting game play would be stable, that is, would the mixed actions of the players converge to an equilibrium?
% We answer this question in the negative for the set of the games which have only completely mixed NE.
We denote the unique NE of the stage game by the tuple $(p^*,q^*)$, where $0 < p^*, q^* < 1$ denote the equilibrium strategies of playing action $1$ by players $1$ and $2$ respectively.
In Theorem~\ref{thm:lastiteratedivergence}, we prove the following statement for any such game (described here informally):
\vspace{1ex}

\begin{center}
\textit{If players $1$ and $2$ use (possibly different) optimal-no-regret, mean-based and monotonic repeated game strategies, then their mixed strategies \textbf{cannot} converge to the NE $(p^*,q^*)$.}
\end{center}

\vspace{1ex}
Our proof technique isolates the ensuing stochasticity in either of the player's realizations as a critical ingredient underlying these \textit{last-iterate oscillations}.
In particular, we prove Theorem~\ref{thm:lastiteratedivergence} via contradiction: suppose, instead, that the sequence of mixed strategies $(\bm{P_t},\bm{Q_t})$ converged to $(p^*,q^*)$, which implies that $\bm{Q_t} \to q^*$.
We show that this would cause sufficient stochasticity by itself to necessitate $\bm{P_t}$ to oscillate with a positive probability \textit{as a fundamental consequence} of no-regret (together with the mean-based and monotonic properties).
The intuition for why stochasticity in realizations is the primary cause of last-iterate oscillations is contained in an elementary ``warm-up'' argument provided in Theorem~\ref{thm:warmup}, which shows that the iterates of player $1$ oscillate even in an idealized scenario in which player $2$ has already converged to his NE strategy, i.e. $q_t = q^*$ for all $t \geq 1$.

The mean-based and monotonic properties described above do not constitute all popular no-regret strategies used in practice; notable exceptions are the family of \textit{optimistic} no-regret algorithms~\citep{rakhlin2013optimization,syrgkanis2015fast,daskalakis2018last} and Online-Mirror-Descent algorithms run with data-adaptive step sizes~\citep{hazan2010extracting,cesa2007improved,erven2011adaptive,rakhlin2013online,rakhlin2013optimization}.
However, we prove that such strategies can be reduced in a natural way to mean-based and monotonic strategies in Section~\ref{sec: beyond_mean_based}, and consequently show that last-iterate oscillations will continue to arise when this broader class of strategies is used.
Our negative result for optimistic strategies in particular highlights a contrast to the \textit{telepathic setting}, in which~\citet{daskalakis2018last} showed that the last iterate (which is deterministic under telepathic dynamics) will converge to NE when both players use optimistic mirror descent strategies.
A complete removal of the mean-based and monotonic assumptions remains an important direction for future work; however, in Section~\ref{sec:conjecture} we provide partial evidence that the monotonicity assumption in particular can be removed.

% {\bf Our techniques in a nutshell:}
% \vidyacomment{In the process of moving this material to relevant main result sections}
% We present the essence of our proof technique in the aforementioned ``warm-up'' idealized case in which player $2$ has already converged to his NE $q^*$.

% \begin{figure}
%     \centering
%     \includegraphics[width=0.5\textwidth]{./Figures/sensitivity_figure}
%     \caption{Depiction of the sensitivity of common no-regret strategies in terms of $f_t(\bm{\widehat{Q}_t})$, as a function of $\bm{\widehat{Q}_t}$ for $t = 10^6$. The no-regret strategies plotted are Online Mirror Descent with the entropy regularizer (also known as multiplicative weights) and log-barrier regularizer, with learning rate $\eta_t = 1/\sqrt{t}$. This choice of learning rate is well-known to yield the optimal regret rate $r = 1/2$. Moreover, for this choice, it can be explicitly verified that a deviation on the order of $1/\sqrt{t}$ in $\bm{\widehat{Q}_t}$ yields a \textit{constant} deviation in the function value $f_t(\bm{\widehat{Q}_t})$ for all $t$.}\label{fig:sensitivity_figure}
% \end{figure}

% \vidyacomment{Right now, this section is only presented for the warm-up case. Do you think we need to say anything about the more challenging adaptive case/martingale arguments here? Maybe one vague sentence on this (e.g. we apply this intuition to the more challenging case in which both players adaptively respond to one another via martingale arguments and essential properties of their time-averages, that ensure sufficient stochasiticity)?}

{\bf Related work:}
While the evolution of the \textit{time-averages} of players' strategies as a consequence of multiple players using no-regret dynamics has been an active topic of study for several decades~\citep{brown1951iterative,robinson1951iterative,foster1997calibrated,fudenberg1998learning,freund1999adaptive,hart2000simple,hart2005adaptive,kalai2005efficient}, the properties of the limiting mixed strategies, or the last-iterates, have only been examined more recently.
This topic has also seen substantial attention in the related setup of \textit{convex-concave min-max optimization}~\citep{daskalakis2018training,mertikopoulos2018optimistic,liang2019interaction,abernethy2019last,lei2020last}, where the primary goal is to attain a pure-strategy NE of a game with a continuous-pure-strategy set through the use of first-order optimization algorithms, e.g. gradient descent-ascent.
This problem has been primarily studied in the \textit{deterministic setting}, corresponding to the aforementioned telepathic dynamics in the game-theoretic setup.
Recently,~\citet{daskalakis2018last} showed that a modification of the multiplicative weights strategy that incorporates \textit{recency bias} succeeds in last-iterate convergence in the game-theoretic setup with telepathic dynamics.
This type of recency bias, commonly called optimism, has also been shown to successfully converge in min-max optimization when applied to the gradient descent/ascent algorithms~\citep{daskalakis2018training,mertikopoulos2018optimistic,liang2019interaction,abernethy2019last,lei2020last}.
Moreover, optimistic algorithms have other notable properties, such as leading to faster convergence rates of the time-average of mixed strategies
 in zero-sum as well as non-zero-sum games~\citep{rakhlin2013optimization,syrgkanis2015fast}.
However, we show in Section~\ref{sec: beyond_mean_based} that when stochastic realization-based feedback is considered, optimistic variants on mean-based strategies \textit{do not} resolve the last-iterate oscillation issue.
In fact, the key phenomena that we outlined above manifest in recency-bias-based strategies as well.
This illustrates that the issue of last-iterate oscillation runs deeper in the traditional repeated-game setting than in the telepathic setting.
We briefly discuss alternative (non-constructive) strategies that could satisfy the last-iterate-convergence property in Section~\ref{sec: concl} --- these strategies are not no-regret, but satisfy a weaker property of ``smoothly calibrated forecasting''~\citep{foster2018smooth}.

%Setup 
%!TEX root = ../main.tex

\section{Setup}\label{sec:setup}

We consider $2 \times 2$ games, i.e. a two player game where both the players have two pure strategies, namely, \textit{action $0$ and action $1$}.
Let the payoff matrices for player $1$ and player $2$ be given by 
\begin{align*}
G := \begin{bmatrix}
G(0,0) & G(0,1) \\
G(1,0) & G(1,1)
\end{bmatrix}
\text{ and }
H := \begin{bmatrix}
H(0,0) & H(0,1) \\
H(1,0) & H(1,1)
\end{bmatrix},
\end{align*}
respectively.
Thus, if player $1$ plays action $i \in \{0,1\}$ and player $2$ plays action $j \in \{0,1\}$, the payoff to player $1$ is given by $G(i,j)$ and the payoff to player $2$ is given by $H(i,j)$.

We denote by the indicator random variables $\bm I$ and $\bm J$ the realizations of the mixed strategies of player $1$ and player $2$, respectively.
We follow the convention of denoting random variables by the bold versions of their corresponding deterministic variables.
Let $p := \bbE [\bm I]$ and $q := \bbE [\bm J]$ be the probabilities with which the two players play action $1$, respectively.
In general, since 
the two players will implement their mixed strategies independently, the random variables $\bm I$ and $\bm J$ will be independent.
Therefore, the expected payoff for player $1$ and player $2$ corresponding to the choice of mixed strategies $(p,q)$ is given by $G(p,q)$ and $H(p,q)$, respectively, where
\begin{align*}
X(p,q) := (1-p)(1-q)X(0,0) + (1-p)q X(0,1) + p(1-q) X(1,0) + pq X(1,1).
\end{align*}
Above, $X$ stands for $G$ or $H$.
In the repeated game setting, $\{\bm{I_t}\}_{t \geq 1}$ and $\{\bm{J_t}\}_{t \geq 1}$ denote the action sequences of the two players and $\{\bm{P_t}\}_{t \geq 1}$ and $\{\bm{Q_t}\}_{t \geq 1}$ denote the mixed strategy sequences of the two players.
We denote by
$${\bm{(I)^t}} := \{\bm{I_s}\}_{s = 1}^t \text{ and  } {\bm{(J)^t}} := \{\bm{J_s}\}_{s = 1}^t$$
the random sequence of actions up to step $t$.
The \emph{empirical averages}, or \textit{time-averages}, of the \emph{actions} of the two players are given by $${\bm{\widehat P_t}} := \frac{1}{t} \sum_{s = 1}^t \bm{I_t} \text{ and } {\bm{\widehat Q_t}} := \frac{1}{t} \sum_{s = 1}^t \bm{J_t},$$ 
respectively.
Similarly, the \emph{empirical averages}, or \textit{time-averages}, of the \emph{mixed strategies} of the two players are given by $${\bm{\overline P_t}} := \frac{1}{t} \sum_{s = 1}^t \bm{P_t} \text{ and } {\bm{\overline Q_t}} := \frac{1}{t} \sum_{s = 1}^t \bm{Q_t},$$
respectively.
General repeated game strategies for player $1$ and player $2$ are given by sequences of functions $\{f_t\}_{t \geq 1}$ and $\{g_t\}_{t \geq 1}$, where for every $t \geq 1$, $f_t, g_t: \{0,1\}^{2(t-1)} \to [0,1]$ map the history up to step $t$ to mixed strategies given by 
$$\bm{P_t} := f_t(\bm{(I)^{t-1}}, \bm{(J)^{t-1}}) \text{ and } \bm{Q_t} := f_t(\bm{(I)^{t-1}}, \bm{(J)^{t-1}})$$ 
for players $1$ and $2$ respectively.
We will refer to these functions $f_t$ and $g_t$ as the \emph{strategy functions} for players $1$ and $2$ at step $t$.
Critically, observe that we are not allowing for telepathy in the updates.
In other words, the history used by player $1$ at step $t$ does not include $\{\bm{Q_s}\}_{s=1}^{t-1}$, and the history used by player $2$ at step $t$ does not include $\{\bm{P_s}\}_{s=1}^{t-1}$.
This is in agreement with the information structure of the traditional repeated game environment.
We now put forward several definitions to identify key sub-classes of these strategies.

\begin{definition}\label{def:selfagnostic}
We say that a repeated game strategy for player $1$ is \emph{self-agnostic} if player $1$ uses only the action sequence of player $2$ to decide her mixed strategy $P_t$ at step $t$.
With an abuse of notation, the strategy function can be replaced by a function $f_t: \{0,1\}^{t-1} \to [0,1]$ such that the mixed strategy for player $1$ at step $t$ is given by $\bm{P_t} = f_t(\bm{(J)^{t-1}})$.
\end{definition}
%
% Note that player $1$ is actually aware of her mixed strategies $\bm{P_1}, \bm{P_2}, \dots, \bm{P_{t-1}}$ at step $t$ since she is aware of her strategy functions $f_1, f_2, \dots, f_{t-1}$ and, for $1 \leq s \leq t-1$, $\bm{P_s}$ can be determined from $f_s$ and $\bm{(J)^{s-1}}$.
Colloquially speaking, the self-agnostic property stipulates that player $1$ is agnostic to the actual realizations of \emph{her own actions} up to that step in the process of choosing her next mixed strategy.

From the point of view of player $1$, we now define \textit{no-regret} strategies, as well as \textit{uniformly no-regret} strategies against an \textit{oblivious opponent}.
The former is precisely the classical definition of consistency first proposed by~\citet{hannan1957approximation}, while the latter is a slightly stronger condition, requiring an effective non-asymptotic guarantee on regret.
We will use the uniform no-regret condition to derive our results.

\begin{definition}
	A self-agnostic repeated game strategy $\{f_t\}_{t \geq 1}$ is said to be \emph{no-regret} if
	\[
		\limsup_{T \to \infty} \frac{1}{T} \left[ \max_{i \in \{0,1\}} \sum_{t=1}^T G(i, J_t) - \sum_{t = 1}^T G(f_t((J)^{t-1}), J_t) \right] \leq 0,
	\]
    for all opponent sequences $\{J_t\}_{t \geq 1}$.
\end{definition}

\begin{definition}
\label{def: uniform_noregret_rate}
	A self-agnostic repeated game strategy $\{f_t\}_{t \geq 1}$ is said to be \emph{uniformly no-regret} if
	\[
        \limsup_{T \to \infty} \max_{\{J_t\}_{t =1}^T} \frac{1}{T} \left[\max_{i \in \{0,1\}} \sum_{t=1}^T G(i, J_t) - \sum_{t = 1}^T G(f_t((J)^{t-1}), J_t) \right] \leq 0.
	\]
	In particular, the strategy achieves a \emph{no-regret rate} of $(r,c)$ if
		\[
        \limsup_{T \to \infty} \max_{\{J_t\}_{t =1}^T} \frac{1}{T^r} \left[\max_{i \in \{0,1\}} \sum_{t=1}^T G(i, J_t) - \sum_{t = 1}^T G(f_t((J)^{t-1}), J_t) \right] \leq c.
	\]
\end{definition}
Observe that all uniformly no-regret strategies are also no-regret.
The converse need not hold --- however, algorithms that are no-regret but not uniformly no-regret are typically contrived examples that are unlikely to be used in practice.

The following properties of uniformly no-regret strategies $\{f_t\}$ can easily be verified:
\begin{enumerate}
    \item If a strategy $\{f_t\}$ satisfies a uniform no-regret rate of $(r,c)$, then it satisfies a no-regret rate of $(r,c')$ for all $c' \geq c$.
    \item If a strategy $\{f_t\}$ satisfies a uniform no-regret rate of $(r,c)$, then it satisfies a no-regret rate of $(r',0)$ for all $r' > r$.
\end{enumerate}
It is well-known (e.g. see~\cite[Chapter 3]{cesa2006prediction}) that, for any finite constant $0 < c < \infty$, the best possible no-regret rate is $r = 1/2$.
Moreover, several commonly used algorithms, such as those in the Online-Mirror-Descent family~\citep{nemirovsky1983problem,shalev2011online}, typically match the optimal no-regret rate for appropriately chosen constant $c$.

We state two additional assumptions satisfied by no-regret strategies commonly encountered in the literature.
We will prove our main results under these assumptions.
\begin{definition}\label{as:meanbased}
The strategy of player $1$ (or $2$) is called \emph{mean-based} if player $1$ uses only the empirical averages of player $2$ (or $1$) as a sufficient statistic to determine her mixed strategy $\bm{P_t}$ at step $t$.
In this case, with another abuse of notation, the strategy function can be replaced by a function $f_t: [0,1] \to [0,1]$, such that the mixed strategy for player $1$ at step $t$ is given by $\bm{P_t} = f_t(\bm{\widehat Q_{t-1}})$.
\end{definition}
For example, all algorithms in the popular Online-Mirror-Descent framework~\citep{nemirovsky1983problem,shalev2011online} satisfy this assumption.
Our results will also apply to variants on the mean-based property that incorporate a recency bias or adaptive step-sizes.
These extensions are discussed in Section~\ref{sec: beyond_mean_based}.

% Another property common to these algorithms is monotonicity defined as follwows: 
\begin{definition}\label{as:monotonic}
The mean-based strategy of player $1$ (see Definitions~\ref{as:meanbased}) is called monotonic if, for every $t \geq 1$, $f_t[\cdot]$ is either \textit{non-increasing} or \textit{non-decreasing} in its argument.
\end{definition}
Note that this assumption \textit{does not} require strict monotonicity, and also does not require the direction of the monotonicity to be the same across rounds.
We can interpret it as a regularity condition that we impose primarily for technical reasons.
% \vidyacomment{Perhaps this is a vague sentence.}
% \mycomment{I agree. But you are right, we need it as a regularity condition. Maybe we can state it as ``We need the monotonicity condition mainly for technical reasons to facilitate a simple proof of our main result. However, we conjecture that ...''}
%\vidyacomment{On second thoughts, I think this sentence may be fine unless you disagree!}
We conjecture that even this regularity condition is not required to show that last-iterate oscillations occur, and discuss partial evidence for this conjecture at length in Section~\ref{sec:conjecture}.

In this paper, we will restrict our attention to $2 \times 2$ games for which none of the NE lie on the boundary of the mixed strategy space, i.e. neither player plays a pure-strategy in any NE.
(It is interesting to note that, although for different reasons, the last-iterate oscillation result shown by~\citet{bailey2018multiplicative} for multiplicative weights in the case of telepathic dynamics also requires the assumption that all the NE lie in the interior of the strategy space.)
% \footnote{Note that, although for different reasons, the last-iterate oscillation result shown by~\citet{bailey2018multiplicative} for multiplicative weights also requires the assumption that all the NE lie in the interior of the strategy space.}.
\citet{phade2019geometry} characterize the NE of all $2 \times 2$ games, and show that all the NE of a $2 \times 2$ game are completely mixed if and only if it is a \emph{competitive game}, defined for completeness below.

\begin{definition}
\label{def: competitive}
A $2 \times 2$ game is said to be competitive if either of the following conditions holds:
	\begin{enumerate}[(a)]
	\item $G(0,0) > G(1,0)$, $G(0,1) < G(1,1)$, $H(0,0) < H(0,1)$, $H(1,0) > H(1,1)$,
	\item $G(0,0) < G(1,0)$, $G(0,1) > G(1,1)$, $H(0,0) > H(0,1)$, $H(1,0) < H(1,1)$.
\end{enumerate}
\end{definition}

Moreover, competitive games have a unique NE $(p^*,q^*)$ which is also its unique correlated equilibrium.
Henceforth, without loss of generality, we will assume that the payoffs satisfy condition (a) above.
Observe that in the special case of zero-sum games, i.e. $H = -G$, the condition (a) would become: $G(0,0) > G(1,0)$, $G(0,1) < G(1,1)$, $G(0,0) > G(0,1)$ and $G(1,0) < G(1,1)$. 
This condition is satisfied by several common zero-sum games, including the \textit{matching pennies} game which we will use as a running example.

% \begin{remark}
% \mycomment{Fix this.}
% The assumption of $0 < p^* < 1$ implies that player $1$ is indifferent between choosing from action $0$ and $1$ when player $2$ is playing his Nash equilibrium strategy $q^*$.
% This is useful for establishing the existence of a randomized opponent sequence against which all choices of strategy would result in the same expected payoff for player $1$.
% The assumption of $0 < q^* < 1$ means that player $2$ actually has stochasticity in his realizations, which is fundamentally important to show the lack of last-iterate convergence for \textit{all} no-regret algorithms.
% \end{remark} 
% \vidyacomment{TODO: mention essence of commented-out remark above in respective results where $0 < p^* < 1$ and $0 < q^* < 1$ are required.}

\begin{remark}\label{remark:competitive}
Note that the set of zero-sum games with unique NE is strictly contained in the set of competitive games. 
The results in \citet{phade2019geometry} imply that, for any competitive game, there exists a corresponding zero-sum game such that the best-response functions of both the players are exactly the same.
This transformation is, in general, non-linear (where a linear transformation is one that is obtained by changing the payoff matrices to be $a_1 G + b_1$ and $a_2 H + b_2$ corresponding to players $1$ and $2$ respectively, for scalars $a_1, b_1, a_2, b_2 \in \bbR$.)
It is easy to see that if two games have the same best-response functions, then they have the same NE. 
However, this does not imply that the players' behavior (whether on average, or day-to-day) will be the same when such games are played repeatedly. 
Appendix~\ref{sec:competitive} illustrates significant qualitative differences in the last-iterate behavior of players playing no-regret strategies for two such ``equivalent'' games.
\end{remark}

We define $R^* := G(p^*, q^*) = G(1,q^*) = G(0,q^*)$ (where the chain of equalities follows because $p^*$ is in the interior and by the definition of a Nash equilibrium).

%Main results AND proofs
%!TEX root = ../main.tex

\section{Main results}\label{sec:results}

In this section, we prove in a series of steps that the limiting mixed strategies as an outcome of both players using optimal, mean-based and monotonic no-regret learning \emph{cannot} converge.
In the process, we will show some fundamental properties about no-regret strategies that hold in more general settings and may be of independent interest.

\subsection{A fundamental sensitivity to fluctuations}\label{sec:fluctuationsensitivity}

We start by proving a condition that any (not necessarily mean-based or monotonic) self-agnostic, no-regret strategy with a rate $r \geq 1/2$ necessarily satisfies.
We show that any such strategy is \emph{fundamentally} sensitive to stochastic fluctuations in the outcomes of player $2$'s strategy, in the sense that such fluctuations will cause player $1$ to deviate from her equilibrium strategy by an additive constant with a constant non-zero 
probability.
More precisely: we show that there must be infinitely many mappings $\{f_t(\cdot)\}_{t \geq 1}$ for player $1$ that deviate (with a constant non-zero probability) by at least a constant value (say $\delta > 0$) from her NE strategy $p^*$ whenever player $2$ deviates from his NE strategy and plays action $1$ (or $0$) for a fraction of steps on the order of $t^{-(1 -r)}$.
When the strategy is mean-based, the deviation of player $2$ for $\mathcal{O}(t^{-(1-r)})$ fraction of steps corresponds to a deviation of $\mathcal{O}(t^{-(1-r)})$ in the empirical average of the actions of player $2$.

\begin{figure}
    \centering
    \begin{subfigure}[b]{0.4\textwidth}
        \includegraphics[width=\textwidth]{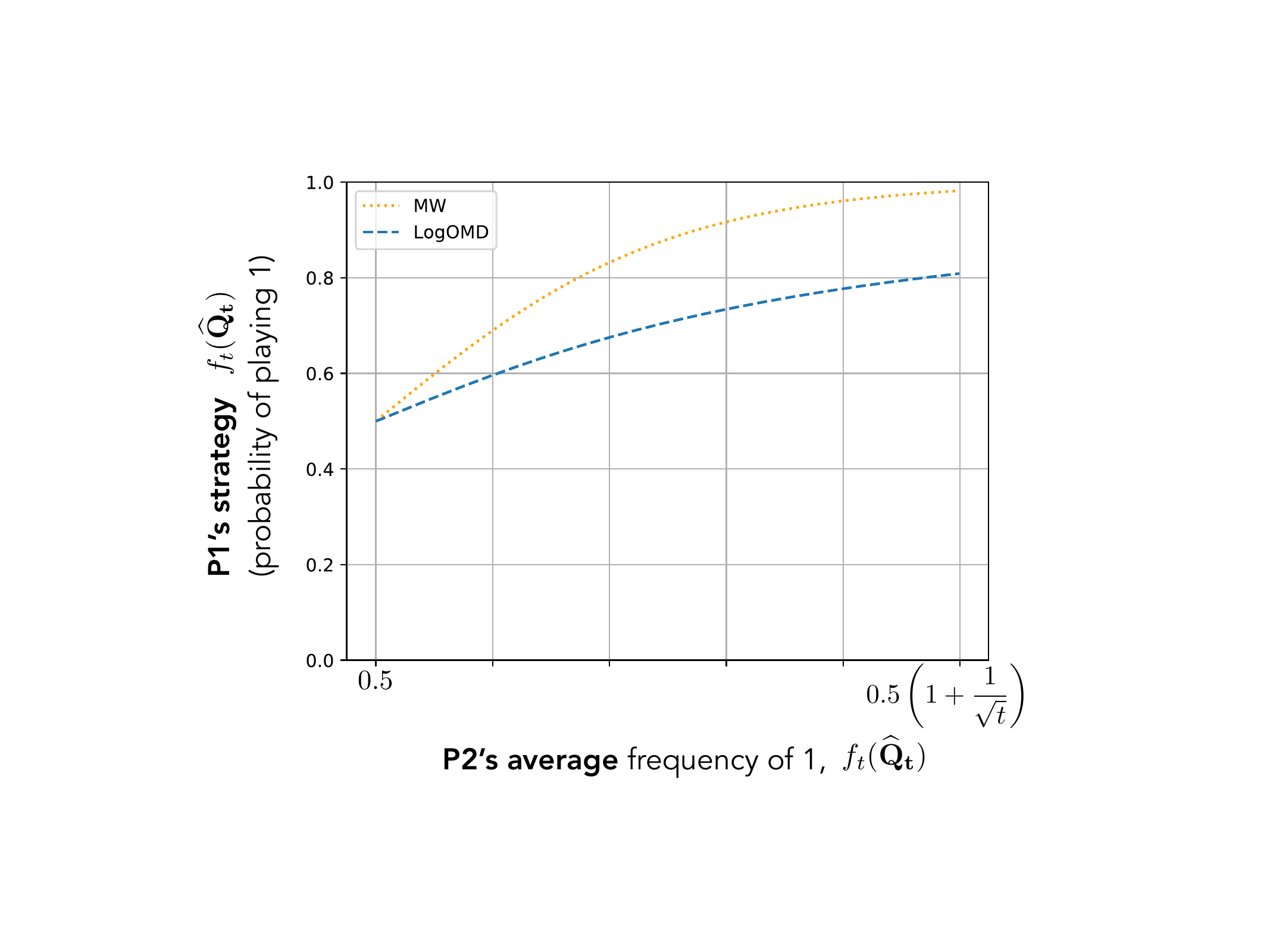}
        \caption{Plot of $f_t(\bm{\widehat{Q}_t})$ v.s. $\bm{\widehat{Q}_t}$ for the no-regret strategies in the Online-Mirror-Descent family with the entropy regularizer (commonly known as multiplicative weights) and log-barrier regularizer, with learning rate $\eta_t = 1/\sqrt{t}$. 
        Here, we set $t = 10^6$.}
        \label{fig:sensitivity_figure}
    \end{subfigure}%
    ~ %add desired spacing between images, e. g. ~, \quad, \qquad etc.
    \begin{subfigure}[b]{0.49\textwidth}
        \includegraphics[width=\textwidth]{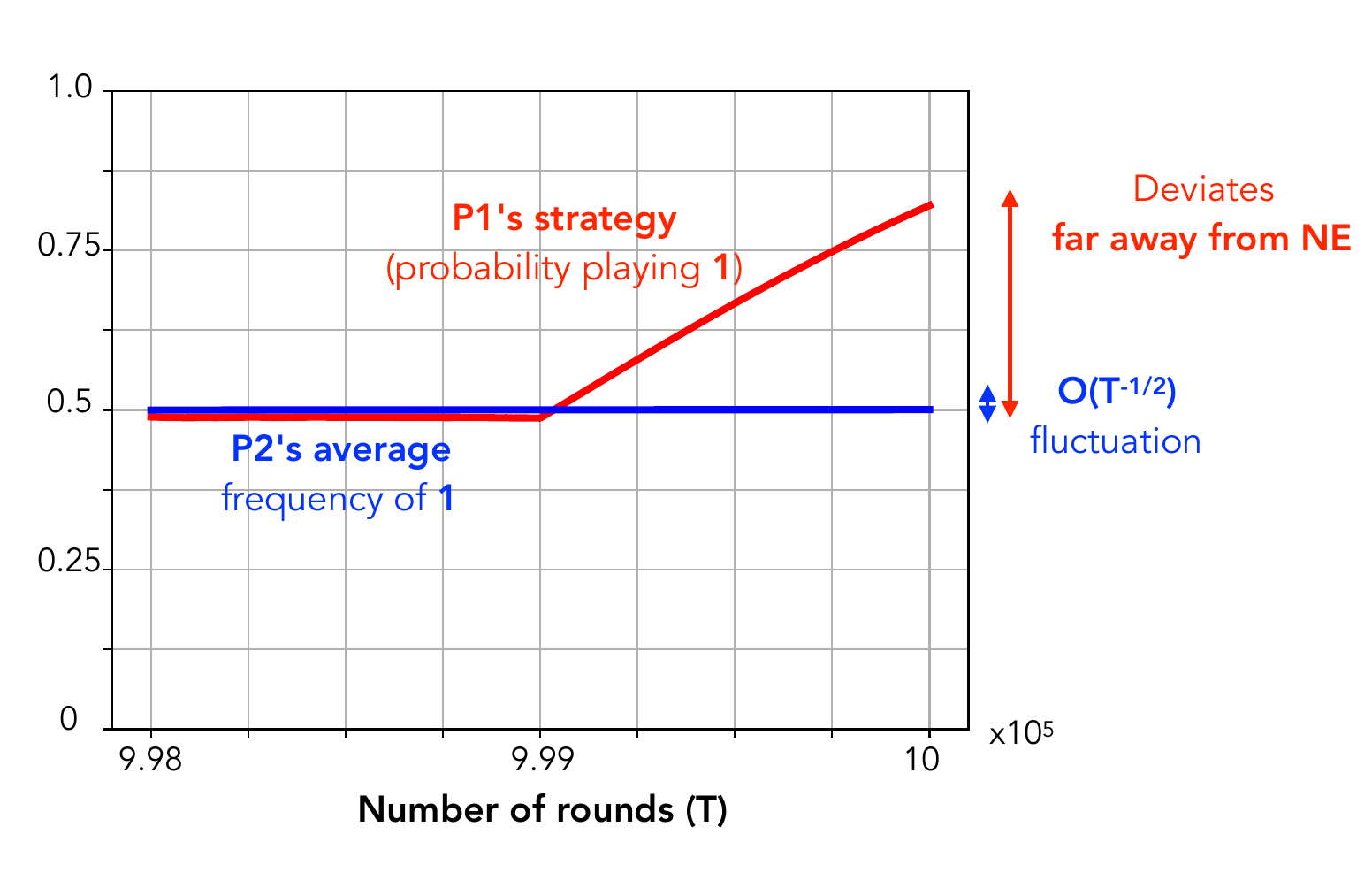}
        \caption{Response of player $1$ against the randomized sequence in Equation~\eqref{eq:randomsequence} (for multiplicative weights).
        The blue line plots the time-averages of player $2$, and the red line plots the iterates of player $1$.
        Since the $y$-axis is on linear scale, the fluctuation in the time-averages of player $2$ (which begins at $s_t := 9.99\times 10^5$) is not visible.}
        \label{fig:sensitivity_figure_proof}
    \end{subfigure}
    \caption{Depiction of the sensitivity of common no-regret strategies in the matching pennies game in terms of $f_t(\bm{\widehat{Q}_t})$, as a function of $\bm{\widehat{Q}_t}$ for $t = 10^6$. Figure~\ref{fig:sensitivity_figure} depicts the extent of sensitivity for two popular no-regret strategies, and Figure~\ref{fig:sensitivity_figure_proof} illustrates the proof strategy.}
    \label{fig:sensitivity}
\end{figure}

Overall, this shows that if we need our no-regret algorithm to have better regret rates, then we have to make the mappings constituting the no-regret algorithm more fluctuation-sensitive.
A depiction of this fluctuation-sensitivity is in Figure~\ref{fig:sensitivity_figure} for the example of the multiplicative weights algorithm, parameterized to have optimal no-regret rate $r = 1/2$, in the matching pennies game.
This property of fluctuation-sensitivity is formally defined and proved below.

% \begin{figure}
%     \centering
%         \includegraphics[width=0.8\textwidth]{./Figures/sensitivity_figure_proof_2}
%     \caption{Response of player $1$, who is using the multiplicative weights algorithm with learning rate $\eta = 1/\sqrt{t}$, against player $2$ who uses the randomized sequence in Equation~\eqref{eq:randomsequence}.
%     The blue line plots the time-averages of player $2$, i.e. $\{\bm{\widehat Q_t}\}_{t \geq 1}$, and the red line plots the iterates of player $1$, i.e. $\{\bm P_t\}_{t \geq 1}$.
%     Note that the $y$-axis of the figure is on linear scale, so the fluctuation in the time-averages of player $2$ is not visible.
%     }\label{fig:sensitivity_figure_proof}
% \end{figure}

\begin{proposition}\label{prop:noregretsensitivity}
For any self-agnostic repeated game strategy 
$\{f_t\}_{t \geq 1}$ used by player $1$ 
that is uniformly no-regret with a rate of $(r,c)$, $1/2 \leq r < 1$
, and any 
$0 < \delta < (1 -p^*)/3$, 
there exists a positive constant $\alpha$ and an infinite sequence of integer tuples $\{(t_k,s_k)\}_{k \geq 1}$ 
such that
\begin{equation}
\label{eq:sensitivitymargin}
    0 < s_k \leq \alpha (t_k)^{r}, \text{ for all } k \geq 1,
\end{equation}
and
\begin{align}
\label{eq:sensitivitycondition}
\mathbb{E}\left[f_{t_k}\left(\bm{(J'(k))^{t_k}}\right)\right] \geq p^* + 2\delta \text{ for all } k \geq 1 ,
\end{align}
where the expectation is over the random sequence 
$\bm{(J'(k))^{t_k}} := \{\bm{J'_s(k)}\}_{s=1}^{t_k}$ 
defined as below:
\begin{align}
\label{eq:randomsequence}
\bm{J'_s(k)} = \begin{cases}
\bm{J^*_s} \text{ i.i.d. } \sim \text{Bernoulli}(q^*), \text{ if } 1 \leq s \leq t_k - s_k, \\
1 \text{ otherwise. }
\end{cases}
\end{align}
Similarly, we also have
\begin{align}
\label{eq:sensitivityconditionopp}
\mathbb{E}\left[f_{t_k}\left(\bm{(J'(k))^{t_k}}\right)\right] \leq p^* - 2\delta \text{ for all } k \geq 1 ,
\end{align}
where the expectation is over the random sequence $\bm{(J''(k))^{t_k}} := \{\bm{J''_s(k)}\}_{s=1}^{t_k}$ defined as below:
\begin{align}\label{eq:randomsequenceopp}
\bm{J''_s(k)} = \begin{cases}
\bm{J^*_s} \text{ i.i.d. } \sim \text{Bernoulli}(q^*), \text{ if } 1 \leq s \leq t_k - s_k, \\
0 \text{ otherwise. }
\end{cases}
\end{align}
\end{proposition}

In particular, if the self-agnostic repeated game strategy $\{f_t\}_{t \geq 1}$ is \textit{optimally} uniformly no-regret i.e. $r = 1/2$, then condition~\ref{eq:sensitivitymargin} would be
\[
   0 < s_k \leq \alpha \sqrt{t_k}, \text{ for all } k \geq 1. 
\]
For this case, Figure~\ref{fig:sensitivity_figure_proof} shows player $1$'s response, when she uses the multiplicative weight algorithm, to an opponent who plays the random sequence defined in Equation~\eqref{eq:randomsequence}.
The game that is being played is matching pennies, which is a zero-sum game with $G(0,0) = G(1,1) = 1$ and $G(0,1) = G(1,0) = -1$.
Player $1$'s strategy, averaged over player $2$'s realizations, is shown to deviate sizably from the NE of the matching pennies game, $p^* = 0.5$.
This case is important because, even if player $2$ remained close to her equilibrium mixed strategy $q^*$ (as is \textit{necessary} in any hypothetical scenario of last-iterate convergence), there is a non-trivial probability of player $2$'s empirical average deviating from $q^*$ by a number on the order of $1/\sqrt{t}$ at step $t$.
The sensitivity in an optimal no-regret strategy to pick deviations of this order will allow us to show the oscillation of player $1$'s mixed strategies in the subsequent Sections~\ref{sec:warmup},~\ref{sec:mainresult},~\ref{sec: beyond_mean_based} and~\ref{sec:conjecture}.
To execute this proof approach, we would use Equation~\eqref{eq:sensitivitycondition} if $f_t(\cdot)$ is monotonically \textit{increasing} in its argument, and Equation~\eqref{eq:sensitivityconditionopp} if $f_t(\cdot)$ is monotonically \textit{decreasing} in its argument.
Henceforth, we will assume that $f_t(\cdot)$ is increasing in its argument without loss of generality.

We conclude this subsection with the proof of Proposition~\ref{prop:noregretsensitivity}.

% \begin{proof}
%This proof actually only requires the NE of player $1$ to be strictly mixed, i.e. $0 < p^* < 1$.
\proof{Proof of Proposition~\ref{prop:noregretsensitivity}.}
Recall that by the competitive $2 \times 2$ game assumption, we have $G(0,1) < G(1,1)$.
Thus, for any $0 \leq p < 1$, we have $G(p,1) < G(1,1)$.
Consider $\{f_t\}_{t \geq 1}$ to be any self-agnostic uniformly no-regret strategy with a no-regret rate of $(r,c)$.
We note that for any $t$, and any sequence $\{J_s\}_{s=1}^t$, we have 
$$\max_{i \in \{0,1\}} \sum_{s=1}^t G(i,J_s) = t \cdot \max\{G(0,\widehat Q_t), G(1,\widehat Q_t)\}.$$
Recall that
$$\widehat Q_t = \frac{1}{t} \sum_{s = 1}^t J_s.$$
Thus, for $c' > c$, there exists a sufficiently large $t_0$ such that for all $t \geq t_0$, we have
\begin{equation}
\label{eq: suff_large_no_regret}
    \frac{1}{t^{r}} \left[t \cdot \max\{G(0,\widehat Q_t), G(1,\widehat Q_t)\} - \sum_{s = 1}^t G(f_s((J)^{s-1}), J_s)) \right] \leq c',
\end{equation}
for any sequence $\{J_s\}_{s = 1}^t$.

Recall that $R^*$ denotes the Nash equilibrium payoff of player $1$.
Because the equilibrium $(p^*,q^*)$ is strictly mixed, we have $R^* = G(p^*,q^*) = G(p^*, 1) < G(1,1)$.
Let $0 < \delta < \frac{(1 - p^*)}{3} = \frac{(G(1,1) - R^*)}{3(G(1,1) - G(0,1))},$ and let $\delta' := \delta(G(1,1) - G(0,1))$.
Note that $0 < \delta' < \frac{(G(1,1) - R^*)}{3}$.
Let $\alpha := \frac{c'}{(G(1,1) - R^* - 3\delta')} > 0$.
For any $t > t_1 := \max\{t_0, \alpha^{1/r} ,(\alpha \delta'/G(1,1))^{-1/r}\}$, let
$t^*(t) := t - \lfloor \alpha t^{r} \rfloor,$ 
where $\lfloor \cdot \rfloor$ is the floor function.
Note that since $t > \alpha^{1/r}$, we have $t^*(t) \geq 1$.
Let $\{\bm{J_s^*}\}$ be an i.i.d. sequence of Bernoulli($q^*$) random variables for $1 \leq s \leq t$.
Let $\bm{\widehat Q_s^*} := \frac{1}{s} \sum_{s' = 1}^s \bm{J_{s'}^*}$ denote the empirical average of this sequence at step $s$.
We now state and prove the following useful lemma.

\begin{lemma}
\label{lem: martingale_prop}
For the sequence $\{\bm{J_s^*}\}_{s \geq 1}$ defined above, and any $t \geq 1$, we have
    \begin{subequations}
    \begin{align}
        \mathbb{E}\left[\sum_{s = 1}^{t} G(f_s(\bm{(J^*)^{s-1}}), \bm{J_s^*}))\right] &= t R^* \text{ and } \label{eq:Gteq}\\
        \mathbb{E}\left[\sum_{s=1}^{t} \bm{J_s^*}\right] &= t q^* \label{eq:Jteq}.
    \end{align}
    \end{subequations}
\end{lemma}

\proof{Proof of Lemma~\ref{lem: martingale_prop}.}
We consider the distribution of \textit{mutually independent} coin tosses,
\begin{align}\label{eq:Jtunif}
    \bm{J_s} \text{ i.i.d. } \sim \text{Bernoulli}(q^*) ,
\end{align}
and denote the expectation of quantities under this probability distribution by $\mathbb{E}[\cdot]$.
Recall that $q^*$ is the Nash equilibrium strategy of player $2$.
By linearity of expectation, it is trivial to show the second statement, i.e.
\begin{align*}
    \mathbb{E}\left[\sum_{s=1}^{t} \bm{J_t}\right] = t q^*.
\end{align*}
To show the first statement, recall that $\bm{J_s} \perp \bm{(J)^{s-1}}$ for all $s \in \{1,\ldots, t\}$ due to mutual independence.
Thus, we use the law of iterated expectations to get
\begin{align*}
    &\mathbb{E}\left[\sum_{s=1}^{t} G(f_s(\bm{(J)^{s-1}}),\bm{J_s})\right]\\
    &= \mathbb{E}\left[\sum_{s=1}^{t} \mathbb{E}\left[f_s(\bm{(J)^{s-1}}) \cdot G(1,\bm J_s) + (1 - f_s(\bm{(J)^{s-1}}))G(0,\bm J_s) \Big{|} \bm{(J)^{s-1}} \right] \right] \\
    &= \mathbb{E}\left[\sum_{s=1}^{t} f_s(\bm{(J)^{s-1}}) \cdot G(1,q^*) + (1 - f_s(\bm{(J)^{s-1}})) \cdot G(0,q^*) \right] \\
    &= t R^*,
\end{align*}
where the last statement follows 
%by statement $1$ in Assumption~\ref{as:2by2assumptions}, noting that 
from $G(0,q^*) = G(1,q^*) = G(p^*,q^*) = R^*$.
This completes the proof.
% \Halmos
\endproof

% See Appendix~\ref{sec: martingaleproof} for the proof of this lemma.
We now use Lemma~\ref{lem: martingale_prop} to prove Proposition~\ref{prop:noregretsensitivity}.
We define the sequence $\{\bm{J_s'}\}_{s=1}^{t}$ as specified in the statement of the proposition.
In other words, we define $\bm{J_s'} = \bm{J^*_s}$ for $1 \leq s \leq t^*(t)$, and $\bm{J_s'} = 1$ for $t^*(t) < s \leq t$.
Then, we can denote the empirical average of this sequence as $\bm{\widehat Q_s'} := \frac{1}{s} \sum_{s' = 1}^s \bm{J_{s'}'}$ for $1 \leq s \leq t$.
We denote
\begin{align}
\label{eq: M_def}
M := \frac{1}{\lfloor \alpha t^{r} \rfloor} \cdot \mathbb{E}\left[\sum_{s = 1}^{\lfloor \alpha t^{r} \rfloor} G\left(f_{t^*(t) + s}\left(\bm{(J')^{t^*(t) + s - 1}}\right),1\right)\right].
\end{align}
From the definition of uniform no-regret, i.e. Equation~\eqref{eq: suff_large_no_regret}, we have
\begin{align*}
    c'
    &\geq \frac{1}{t^{r}} \cdot \mathbb{E} \left[t \max\{G(0,\bm{\widehat Q_t'}), G(1,\bm{\widehat Q_t'})\} - \sum_{s = 1}^{t} G(f_s(\bm{(J')^{s-1}}), \bm{J_s'})) \right] \\
    &\geq \frac{1}{t^{r}} \cdot \mathbb{E} \left[t \cdot G(1,\bm{\widehat Q_t'}) - \sum_{s = 1}^{t} G(f_s(\bm{(J')^{s-1}}), \bm{J_s'})) \right] \\
    &= \frac{1}{t^{r}} \cdot \left[ t^*(t) \cdot R^* + \lfloor \alpha t^{r} \rfloor \cdot G(1,1) - t^*(t) \cdot R^* - \lfloor \alpha t^{r} \rfloor \cdot M \right] \\
    &\geq (G(1,1) - M) \alpha -  G(1,1) t^{-r}.
\end{align*}
Using the fact that $\alpha := \frac{c'}{(G(1,1) - R^*_1 - 3\delta')}$ and $t > \left(\frac{\alpha \delta'}{G(1,1)}\right)^{-1/r}$, we get

\begin{align*}
    M \geq G(1,1) - \frac{c'}{\alpha} - \frac{G(1,1) \cdot t^{-r}}{\alpha}
    = R^* + 3\delta' - \frac{G(1,1) \cdot t^{-r}}{\alpha}
    \geq R^* + 2\delta' .
\end{align*}
Now, using linearity of expectation, and linearity of the payoff function $G(p,1)$ in the argument $p$, we get
\begin{align*}
    M &= G(\overline{f}, 1) \text{ where } \\
    \overline{f} &:= \frac{1}{\lfloor \alpha t^r \rfloor}  \sum_{s = 1}^{\lfloor \alpha t^r \rfloor} \mathbb{E}\left[f_{t^*(t) + s}\left(\bm{(J')^{t^*(t) + s - 1}}\right)\right] .
\end{align*}
Now, since $G(1,1) > G(0,1)$, we note that $G(p,1) = G(0,1) + (G(1,1) - G(0,1)) p$ is an increasing function in $p$ and so we get
\begin{align*}
\overline{f} \geq p^* + \frac{2\delta'}{G(1,1) - G(0,1)} = p^* + 2 \delta, 
\end{align*}
from the above inequality on $M$.
Thus, there exists $s(t)$ such that $1 \leq s(t) \leq \lfloor \alpha t^{r} \rfloor$ and 
\begin{align}\label{eq:sensitivitycondition2}
    \bbE \left[ f_{t^*(t) + s(t)}\left(\bm{(J')^{t^*(t) + s(t) - 1}}\right) \right] \geq p^* + 2\delta.
\end{align}
To write this in the language of Equation~\eqref{eq:sensitivitycondition}, we observe that
\begin{align*}
    0 < \frac{s(t)}{t^*(t) + s(t)} \leq \frac{\lfloor \alpha t^{r} \rfloor}{t^*(t) + \lfloor \alpha t^{r} \rfloor} \leq \frac{\alpha t^{r}}{t} = \alpha t^{r-1} \leq \alpha (t^*(t) + s(t))^{r-1}, 
\end{align*}
and therefore we get,
\begin{align}\label{eq:stinequality}
    s(t) \leq \alpha (t^*(t) + s(t))^{r} .
\end{align}

We note that $t^*(t) \to \infty$ as $t \to \infty$, and hence we can define an infinite sequence of integer tuples $\{t_k, s_k\}_{k \geq 1}$ (where we have defined $t_k = t^*(t_1 + k) + s_k$ and $s_k = s(t_1 + k)$ as above) such that $0 < s_k \leq \alpha (t_k)^{r}$ and
\begin{align}
\label{eq: twisted_response_exp}
\mathbb{E}\left[f_{t_k}\left(\bm{(J'(k))^{t_k}}\right)\right] \geq p^* + 2\delta \text{ for all } k = 1,2, \ldots
\end{align}
This is precisely the statement in Equation~\eqref{eq:sensitivitycondition}, and completes the proof of Proposition~\ref{prop:noregretsensitivity}.
% \Halmos
\endproof
% \end{proof}

\subsection{Warmup: Last-iterate oscillation when opponent is already at equilibrium}\label{sec:warmup}

Equation~\eqref{eq:sensitivitycondition} highlights a critical property of any self-agnostic uniformly no-regret algorithm with a regret rate of $(r,c)$: 
it needs to be sufficiently sensitive to small perturbations on the order of $t^r$ in the opponent's strategy.
We can concretize this property to show that the \emph{last-iterate oscillates} (i.e. does not converge almost surely) when both players use optimal no-regret strategies (i.e. $r = 1/2$) that are mean-based (Definition~\ref{as:meanbased}) and monotonic (Definition~\ref{as:monotonic}).
Recall that under the mean-based assumption, player $1$'s strategy functions are given by $f_t(\bm{(J)^{t-1}}) := f_t(\bm{\widehat{Q}_{t-1}}) \text{ for all } t \geq 1$.

\begin{figure}
    \centering
        \includegraphics[width=0.5\textwidth]{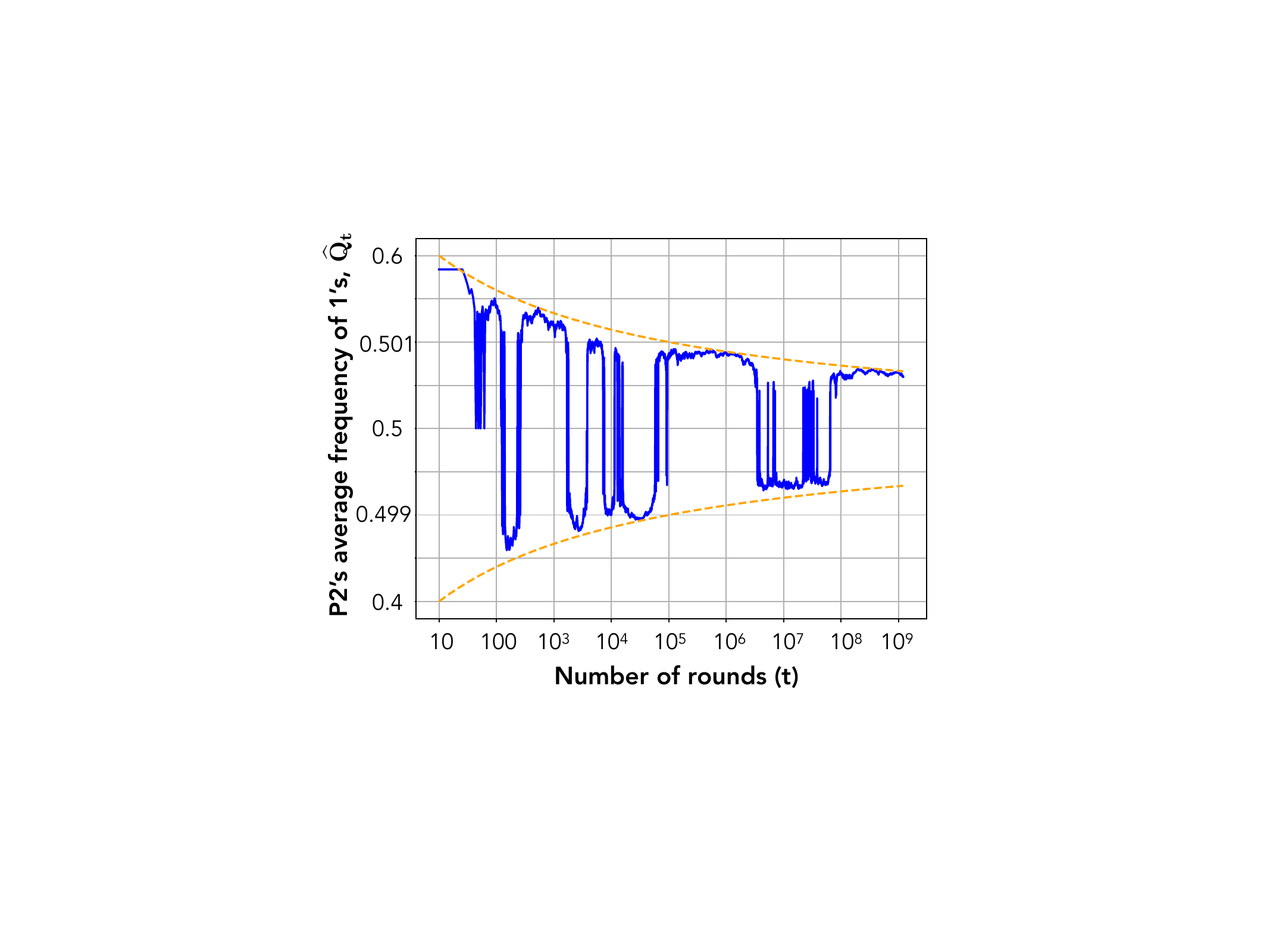}
    \caption{Depiction of constant fluctuations of player $2$'s time-averages, i.e. $\bm{\widehat{Q}_t}$, as a function of $t$, when player $2$ plays $\bm{Q_t} \text{ i.i.d } \sim q^*$.}\label{fig:average_iterates_figure}
\end{figure}

In this section, we state and prove a warm-up result that clearly illustrates how the stochasticity in realizations \textit{alone} can lead to last-iterate oscillation.
We consider the idealized case in which player $2$ is playing his equilibrium strategy at all steps, i.e. $\{\bm J_t\}_{t \geq 1} \text{ i.i.d} \sim \text{Bernoulli}(q^*)$.
Remarkably, we show that even this simple case necessitates the limiting mixed strategy of player $1$ to diverge!
(Interestingly, this is in stark contrast to the setting of telepathic dynamics, in which a simple algorithm like multiplicative weights used in the matching pennies game would trivially lead player $1$ to converge to the NE strategy, $p^* = 1/2$.)
The central reason for this divergence is as follows: if player $2$ is playing the mixed NE $q^*$ at all his steps, then the time-averages of his realized actions will fluctuate on the order of $t^{-1/2}$ infinitely often as well, leading to a $\delta$-deviation of player $1$ from her NE strategy $p^*$ with a positive probability via Proposition~\ref{prop:noregretsensitivity}.
% In fact, this happens with a fixed positive probability.
Figure~\ref{fig:average_iterates_figure} (which should remind the reader of the fluctuations of a symmetric random walk) depicts these recurring fluctuations on the order of $t^{-1/2}$ of player $2$'s time-averaged actions for a typical realization of player $2$ playing mixed strategy $q^* = 1/2$ at all steps.
The ensuing last-iterate oscillation is stated and proved below.
% \vidyacomment{For now I'm keeping all text here as-is, but we will return to discussing the discrepancy between a fixed strategy by player $2$ and an adaptive one in subsequent sections.}

% Next, we specialize to the case of optimal-no-regret strategies, i.e. $r = 1/2$.

% Putting these observations together, we see that optimal no-regret strategies \textit{need} to be sensitive to the fluctuations of player $2$ infinitely often.
% Moreover, these fluctuations happen infinitely often as well, as a consequence of the stochasticity in player $2$'s realizations.
% These constitute the \emph{key phenomena that underlie last-iterate oscillation for such a broad class of optimal no-regret strategies}.

\begin{theorem}\label{thm:warmup}
    Let player $2$'s strategy $\{\bm{J_t}\}_{t \geq 1}$ be an i.i.d. sequence of Bernoulli($q^*$) random variables.
    Then, any mean-based and monotonic repeated game strategy $\{f_t\}_{t \geq 1}$ that has a regret rate of $(1/2,c)$ causes player $1$'s last iterate to diverge in probability, i.e. there exist positive constants $(\delta,\epsilon)$ such that
    \begin{align}
    \label{eq: liminf_prob_dev_response}
        {\lim \sup}_{t \to \infty} \bbP\left[|\bm{P_{t}} - p^*| \geq \delta \right] \geq \epsilon .
    \end{align}
\end{theorem}
% The proof of Theorem~\ref{thm:warmup} constitutes an elementary application of Markov's inequality, and a change-of-measure argument on the probability mass functions of two binomial random variables.
The proof of Theorem~\ref{thm:warmup} constitutes an application of Markov's inequality and the central limit theorem, and is provided below.
We also prove a version of this theorem without the monotonicity assumption in Proposition~\ref{prop:warmup_ext} in Appendix~\ref{sec:fixedconvergent}.
%!TEX root = ../main.tex
% \begin{proof}{}
\proof{Proof of Theorem~\ref{thm:warmup}.}
We start by defining some notation pertinent to mean-based strategies.
Let $\bm{Z_t} \sim \text{Binomial}(t, q^*)$, for $t \geq 1$.
Let $\bm{Z''_{t, s}} \sim \bm{Z_{t-s}} + s$ and $\bm{Q''_{t, s}} = \bm{Z''_{t, s}}/t$, for $0 \leq s \leq t$, $t \geq 1$.
Let $\{(t_k, s_k)\}_{k \geq 1}$ be an infinite sequence and $0 < \delta < (1 - p^*)/3$ as in Proposition~\ref{prop:noregretsensitivity}.
Thus, for every $k \geq 1$, we have
\begin{align*}
    \bm{Q''_{t_k, s_k}} \overset{d}{=} \frac{1}{t} \sum_{s = 1}^t \bm{J'_t(k)},
\end{align*}
where $\overset{d}{=}$ denotes that the two random variables are identical in distribution, and $\bm{J_t'(k)}$ are random variables as defined in Proposition~\ref{prop:noregretsensitivity}.
We thus have
\begin{equation}
    0 < s_k \leq \alpha (t_k)^{1/2}, \text{ for all } k \geq 1,
\end{equation}
and
\begin{align}
\mathbb{E}\left[f_{t_k}\left(\bm{Q''_{t_k, s_k}}\right)\right] \geq p^* + 2\delta, \text{ for all } k \geq 1 ,
\end{align}

Consider the random variable $\bm{Y} := 1 - f_{t_k}(\bm{\widehat{Q}''_{t_k, s_k}})$.
Since the range of $f_t$ is always $[0,1]$, we have $\bm{Y} \geq 0$.
Thus, we have $\mathbb{E}[\bm{Y}] \leq 1 - (p^* + 2\delta) = (1 - p^*) - 2\delta$.
By Markov's inequality, we have
\[
 \bbP\left(\bm{Y} \geq (1 - p^*) - \delta\right) \leq \frac{(1 - p^*) - 2\delta}{(1 - p^*) - \delta}.   
\]
Thus, we get
\begin{equation}
\label{eq: prob_Qt_lowerbdd}
     \bbP\left(f_{t_k}\left(\bm{\widehat{Q}''_{t_k, s_k}}\right) > p^* + \delta\right) \geq \epsilon_0,
\end{equation}
where we define $\epsilon_0 := {\delta}/{((1 - p^*) - \delta)}$.
Note that $0 < \epsilon_0 < 1/2$ (because we had defined $0 < \delta < (1 - p^*)/3$).
By the central limit theorem, we know that 
\[
     \lim_{t \to \infty} \bbP\left(\bm{\widehat{Q}_{t}} > q^* + \frac{\gamma}{\sqrt{t}}  \right) = 1 - \Phi\left(\frac{\gamma}{\sqrt{q^*(1-q^*)}}\right),
 \] 
where $\Phi(\cdot)$ is the cumulative distribution function of the standard normal distribution.
Now, note that the function $\Phi(\cdot) : \bbR \to (0,1)$ in continuous, strictly increasing.
Moreover, we have $\Phi(0) = 1/2$ and $\lim_{x \to \infty} \Phi(x) = 1$.
Hence, there exists $\gamma_0 > 0$ such that $1 - \Phi(\gamma_0/\sqrt{q^*(1-q^*)}) = \epsilon_0/4 < 1/2$.
Consequently, there exists $T_1'(\epsilon_0) > 1$ such that
\[
     \bbP\left(\bm{\widehat{Q}_{t}} > q^* + \frac{\gamma_0}{\sqrt{t}}  \right) \leq (1 - \phi(\gamma_0)) + \frac{\epsilon_0}{4} \leq \frac{\epsilon_0}{2},
 \] 
for all $t \geq T_1'(\epsilon_0)$.
Now, observe that
\[
    t_k\bm{\widehat{Q}''_{t_k, s_k}} \overset{d}{=} (t_k - s_k) \bm{\widehat{Q}_{t_k - s_k}} + s_k.
\]
Since $t_k - s_k \to \infty$ as $k \to \infty$,
there exists a $k_1 > 1$ such that 
 \[
     \bbP \left(t_k \bm{\widehat{Q}''_{t_k, s_k} } > q^*(t_k - s_k) + s_k + \gamma_0 \sqrt{t_k - s_k}  \right) \leq \frac{\epsilon_0}{2},
 \]
 for all $k \geq k_1$.
 Using the bound $s_k \leq \alpha \sqrt{t_k}$, we get
 \begin{equation}
 \label{eq: prob_tail_lowerbdd}
     \bbP \left(\bm{Z''_{t_k, s_k}} > q^* \cdot t_k + \beta \sqrt{t_k} \right) \leq \frac{\epsilon_0}{2}
 \end{equation}
for all $k \geq k_1$, where $\beta := \alpha + \gamma_0$.
From the union bound and Equations~\eqref{eq: prob_Qt_lowerbdd} and \eqref{eq: prob_tail_lowerbdd}, we get
\begin{equation}
\label{eq: prob_markov_tailcut_bdd}
    \bbP \left(f_{t_k} \left(\frac{\bm{{Z}''_{t_k, s_k}}}{t_k} \right) \geq p^* + \delta, \bm{Z''_{t_k, s_k}} \leq q^* \cdot t_k + \beta \sqrt{t_k} \right) \geq \frac{\epsilon_0}{2}.
\end{equation}
Now, we use the monotonicity assumption (Definition~\ref{as:monotonic}) to show that this requires
\begin{align}\label{eq:deterministiccondition}
f_{t_k}\left(q^* + \frac{\beta}{\sqrt{t_k}}\right) \geq p^* + \delta \text{ for all } k \geq 1 .
\end{align}
This follows because, if Equation~\eqref{eq:deterministiccondition} were \textit{not} true, we would get $f_{t_k}(z/t_k) < p^* + \delta$ for all $z \leq q^* + \beta/\sqrt{t_k}$ by Definition~\ref{as:monotonic}, which would violate the statement of Equation~\eqref{eq: prob_markov_tailcut_bdd}.
Applying the monotonicity assumption again gives us
\begin{align}\label{eq:noregretsensitivity2}
f_{t_k}(x) \geq p^* + \delta \text{ for all } x \geq q^* + \frac{\beta}{\sqrt{t_k}} \text{ and for all } k \geq 1 .
\end{align}
Recall that $\bm{P}_{t_k}$ refers to the individual iterates, and $\widehat{\bm{Q}}_{t_k}$ refers to the time-average of iterates.
The proof of Theorem~\ref{thm:warmup} is now completed by another elementary application of the central limit theorem, as detailed below:
\begin{align*}
\bbP\left[\bm{P}_{t_k} \geq p^* + \delta \right] &= \bbP\left[f_{t_k}(\widehat{\bm{Q}}_{t_k-1}) \geq p^* + \delta \right] \\
&\stackrel{(\mathsf{i})}{\geq} \bbP\left[\widehat{\bm{Q}}_{t_k-1} \geq q^* + \frac{\beta}{\sqrt{t_k}}\right] \\
&\stackrel{(\mathsf{ii})}{\geq} \frac{1}{2} \cdot \bbP\left[W \geq \frac{\beta}{\sqrt{q^*(1 - q^*)}} \right] \text{ where } W \sim \mathcal{N}(0,1) \\
&= \frac{1}{2} \text{erfc}\left(\frac{\beta}{\sqrt{q^*(1 - q^*)}}\right) := \epsilon(\delta) > 0 \text{ for any } \delta < 1/4 .
\end{align*}
Above, inequality $(\mathsf{i})$ follows from Equation~\eqref{eq:noregretsensitivity2} and inequality $(\mathsf{ii})$ follows from the central limit theorem, where $1/2 < 1$ is chosen as an arbitrary constant.
This completes the proof.
% \Halmos
\endproof
% \end{proof}
%!TEX root = ../main.tex

\subsection{Main result: Last-iterate oscillation when \textit{both} players use no-regret}\label{sec:mainresult}

The case for which player $2$ is \textit{already} at equilibrium serves to isolate the ramifications of the stochasticity in realizations.
While the day-to-day behavior of players is clearly quite different when both players are using no-regret strategies (see Figures~\ref{fig:MPoptimalfixedNE} and~\ref{fig:MPoptimalstochastic} in Section~\ref{sec:simulations} for a comparison), we show below that the inherent stochasticity in realizations continues to be the dominating factor that causes last-iterate oscillations.
Our main result is stated below.
\begin{theorem}\label{thm:lastiteratedivergence}
If both players $1$ and $2$ use optimal no-regret repeated game strategies that are mean-based (Definition~\ref{as:meanbased}) and monotonic (Definition~\ref{as:monotonic}), then the pair of mixed strategies of both the players $(\bm{P_t}, \bm{Q_t})$ cannot converge to the NE $(p^*,q^*)$ almost surely.
\end{theorem}
The proof of Theorem~\ref{thm:lastiteratedivergence} mirrors the proof of the simpler ``warm-up'' case (Theorem~\ref{thm:warmup}) by taking a \textit{proof-by-contradiction} approach: Suppose, instead, that both players \textit{did} converge almost surely.
Then player $2$ would have to converge almost surely by definition, and we show that in this scenario player $1$ cannot converge because player $2$'s realizations are effectively too stochastic to allow it.
The primary challenge is in showing this sufficient stochasticity in player $2$'s realizations; after all, player $2$'s realizations are highly dependent on past outcomes of player $1$.
To do this, we use the martingale central-limit-theorem~\citep{hall2014martingale} together with the well-known property of convergence of the \textit{time-average} of the mixed strategies to NE~\citep{freund1999adaptive,roughgarden2016twenty}.
The full proof is provided below.
%!TEX root = ../main.tex

% \begin{proof}
\proof{Proof of Theorem~\ref{thm:lastiteratedivergence}.}
We will use the approach of proof-by-contradiction.
Suppose that we in fact have $(\bm{P}_t \to p^*, \bm{Q}_t \to q^*) \text{ a.s.}$.
Note that this trivially implies that $\bm{Q}_t \to q^* \text{ a.s.}$
To complete the proof by contradiction, we will show that if $\bm{Q}_t \to q^* \text { a.s.}$, then we need to have
\begin{align*}
\lim_{t \to \infty} \bbP\left[\bm{P}_t \geq p^* + \delta \right] \geq \epsilon(\delta) > 0 \text{ for any } 0 < \delta < \frac{(1 - p^*)}{3}. 
\end{align*}
Here $\epsilon(\delta)$ is as defined in the proof of Theorem~\ref{thm:warmup}.
Our first critical ingredient is the following claim, which implies sufficient stochasticity in the iterates under the assumed almost sure convergence of the mixed strategies of player $2$.
% We prove this claim in Appendix~\ref{sec: claim_as_fracconv}.
\begin{claim}
\label{clm: conv_as_frac}
$\bm{Q}_t \to q^* \text{ a.s. }$ implies that
\begin{align}\label{eq:continuumclosetoNE}
{\lim \inf}_{k \to \infty} \frac{1}{t_k} \Big{|} \left\{s \leq t_k: \frac{q^*}{2} \leq \bm{Q}_s \leq \frac{q^* + 1}{2}\right\} \Big{|} = 1 \text{ almost surely, }
\end{align}
for any deterministic sub-sequence $\{t_k\}_{k \geq 1}$.
\end{claim}
Equation~\eqref{eq:continuumclosetoNE} is reminiscent of a convergence condition that is used by~\citet{foster2018smooth} for ``smoothly calibrated'' strategies.
We prove Claim~\ref{clm: conv_as_frac} below.
\proof{Proof of Claim~\ref{clm: conv_as_frac}.}
Let $\Omega$ denote the $\sigma$-algebra induced by $\{\bm{P}_t,\bm{Q}_t\}_{t \geq 1}$ and the strategies $\{f_t(\cdot),g_t(\cdot)\}_{t \geq 1}$.
Let $\omega \in \Omega$ denote any realization of this, and $(\bm{P}_t(\omega),\bm{Q}_t(\omega))$ denote the corresponding mixed strategies at time $t$.
Then, $\bm{Q}_t \to q^*$ a.s. implies that
\begin{align*}
\bbP\left(\omega \in \Omega: \lim_{t \to \infty} \bm{Q}_t(\omega) = q^* \right) = 1 .
\end{align*}
Thus, we get
\begin{align*}
\bbP\left(\omega \in \Omega: \lim_{k \to \infty} \bm{Q}_{t_k}(\omega) = q^* \right) = 1
\end{align*}
for any subsequence $\{t_k\}_{k \geq 1}$.
This implies, by the definition of a limit, that for every $\omega$ such that $\lim_{t \to \infty} \bm{Q}_{t}(\omega) = q^*$, there exists finite $t_0(\omega)$ such that
\begin{align*}
\bm{Q}_{t}(\omega) \in \left[\frac{q^*}{2},\frac{q^* + 1}{2} \right] \text{ for all } t \geq t_0(\omega) .
\end{align*}
Let $k_0(\omega) := \min\{k: t_k \geq t_0(\omega)\}$.
This directly gives us
\begin{align*}
{\lim \inf}_{k \to \infty} \frac{1}{t_k} \Big{|} \left\{s \leq t_k: \frac{q^*}{2} \leq \bm{Q}_s(\omega) \leq \frac{q^* + 1}{2}\right\} \Big{|} \geq {\lim \inf}_{k \to \infty} \frac{t_k - t_{k_0(\omega)}}{t_k} = 1 .
\end{align*}
Thus, we get
\begin{align*}
\bbP\left(\omega \in \Omega: {\lim \inf}_{k \to \infty} \frac{1}{t_k} \Big{|} \left\{s \leq t_k: \frac{q^*}{2} \leq \bm{Q}_s(\omega) \leq \frac{q^* + 1}{2}\right\} \Big{|} = 1 \right) = 1 ,
\end{align*}
which completes the proof of the claim.
% \Halmos
\endproof

As a direct consequence of the claim, we have
\begin{align}\label{eq:variancebound}
{\lim \inf}_{t \to \infty} \frac{1}{t} \sum_{s=1}^t \bm{Q}_s(1 - \bm{Q}_s) \geq \min\left\{\frac{q^*(2 - q^*)}{4},\frac{(1 + q^*)(1 - q^*)}{4} \right\} := \gamma(q^*) \text{ a.s. } 
\end{align}
We consider the filtration $\{\mathcal{F}_t := (\bm{I}_s,\bm{J}_s)_{s=1}^{t-1}\}_{t \geq 1}$, and the stochastic process $\{\bm{D}_t := \bm{Z}_t - \bm{Z}_{t-1} - \bm{Q}_t\}_{t \geq 1}$.
Recall that we had defined $\bm{Z}_t := \sum_{s=1}^t \bm{J}_t$, where $\bm{J}_t \sim \text{Ber}(\bm{Q}_t)$.
We observe the following properties of the stochastic process $\{\bm{D}_t\}_{t \geq 1}$:
\begin{enumerate}
\item $\bm{D}_t$ is a martingale difference sequence with respect to the filtration $\mathcal{F}_{t-1}$. 
This is a direct consequence of $\bm{Q}_t$ being a \textit{deterministic} function of $\mathcal{F}_{t-1}$.
\item $|\bm{D}_t| \leq 1$.
\item Note that because $0 < q^* < 1$, we have $\gamma(q^*) > 0$.
Therefore, the sum of conditional variances diverges to $\infty$.
In other words, we have
\begin{align*}
\sigma_t^2 := \bbE\left[\bm{D}_t^2 | \mathcal{F}_{t-1}\right] &= \bbE\left[(\bm{J}_t - \bm{Q}_t)^2 | \mathcal{F}_{t-1}\right] \\
&= \bm{Q}_t(1 - \bm{Q}_t)
\end{align*}
and from Equation~\eqref{eq:variancebound} it is clear that ${\lim \inf}_{t \to \infty} \sum_{s=1}^t \sigma_s^2 = \infty$ a.s.
\end{enumerate}
As a consequence of these properties, we can apply the martingale central limit theorem~\citep{hall2014martingale} in the following form:

\begin{theorem}[Martingale CLT,~\citep{hall2014martingale}]\label{thm:martingaleCLT}
Let $\{\bm{D}_t\}_{t \geq 1}$ be a martingale difference sequence with respect to the filtration $\mathcal{F}_{t-1}$ such that $|\bm{D}_t| \leq 1$.
Further, let $\{\sigma_t^2 := \bbE\left[\bm{D}_t^2 | \mathcal{F}_{t-1}\right]\}_{t \geq 1}$ denote the conditional variance sequence with the property that $\sum_{t=1}^{\infty} \sigma_t^2$ diverges to infinity almost surely.
Then, we have
\begin{align}\label{eq:martingaleCLT}
\frac{\sum_{s=1}^t \bm{D}_s}{\sqrt{\sum_{s=1}^t \sigma_s^2}} \to \bm{Z} \sim \mathcal{N}(0,1) .
\end{align}
\end{theorem}
Note that Equation~\eqref{eq:martingaleCLT} directly implies
\begin{align}\label{eq:martingaleCLTsubsequence}
\frac{\sum_{s=1}^{t_k} \bm{D}_s}{\sqrt{\sum_{s=1}^{t_k} \sigma_s^2}} \to \bm{Z} \sim \mathcal{N}(0,1) .
\end{align}
for any deterministic subsequence $\{t_k\}_{k \geq 1}$.
Further, note that for any $k \geq 1$, we have
\begin{align*}
\frac{\sum_{s=1}^{t_k} \bm{D}_s}{\sqrt{t_k}} = \frac{1}{\sqrt{t_k}}\left( \bm{Z}_{t_k} - \sum_{s=1}^{t_k} \bm{Q}_s \right) = \sqrt{t_k} (\widehat{\bm{Q}}_{t_k} - \overline{\bm{Q}}_{t_k}).
\end{align*}
Consequently, we have for any $\beta' > 0$, 
\begin{align*}
\lim_{k \to \infty} \bbP\left(\sqrt{t_k} (\widehat{\bm{Q}}_{t_k} - \overline{\bm{Q}}_{t_k}) \geq \beta' \right) &= \lim_{k \to \infty} \bbP\left( \frac{\sum_{s=1}^{t_k} \bm{D}_s}{\sqrt{t_k}} \geq \beta' \right) \\
&\stackrel{\mathsf{(i)}}{\geq} \lim_{k \to \infty} \bbP\left(\frac{\sum_{s=1}^{t_k} \bm{D}_s}{\sqrt{\sum_{s=1}^{t_k} \bm{Q}_s(1 - \bm{Q}_s)}}\geq \frac{\beta'}{\gamma(q^*)} \right) \\
&\stackrel{\mathsf{(ii)}}{\geq} \text{erfc}\left(\frac{\beta'}{\gamma(q^*)}\right) > 0 .
\end{align*}
%
% where $c := \frac{16\beta'}{3}$.
Above, inequality $(\mathsf{i})$ uses Equation~\eqref{eq:continuumclosetoNE}, and inequality $(\mathsf{ii})$ uses the martingale CLT on the subsequence $\{t_k\}_{k \geq 1}$ (as described in Equation~\eqref{eq:martingaleCLTsubsequence}.

Our next critical observation is a fundamental and well-known~\citep{freund1999adaptive,roughgarden2016twenty} property of the \textit{average} iterates arising as a consequence of \textit{both} players using optimal no-regret algorithms.
Specifically,~\cite{freund1999adaptive} showed that the time-averaged \emph{payoff} of players in a zero-sum game converges to the unique Nash equilibrium payoff, and~\cite{roughgarden2016twenty} showed that the average iterates in a general non-zero-sum game converge to the polytope of \emph{coarse correlated equilibria}.
The lemma below states that, as a result, the average iterates in a competitive game converge to its unique Nash equilibrium.
% We state this property as a lemma.
\begin{lemma}[\citep{roughgarden2016twenty}]\label{lem:timeave}
Suppose players $1$ and $2$ both use optimal no-regret strategies in a competitive game.
Then, there exists a universal constant $C > 0$ such that
\begin{align}\label{eq:timeave}
|\overline{\bm{Q}}_t - q^*| \leq \frac{C}{\sqrt{t}} \text{ pointwise, for all } t \geq 1.
\end{align}
\end{lemma}
Lemma~\ref{lem:timeave}, which follows as a corollary of~\cite{roughgarden2016twenty}, is proved in Appendix~\ref{sec:timeave} for completeness.
Selecting $\beta' := \beta + C$ for any $\beta > 0$ (where $C$ is the constant of choice in Lemma~\ref{lem:timeave}), we get
\begin{align}\label{eq:effectiveCLT}
\lim_{k \to \infty} \bbP\left(\sqrt{t_k} (\widehat{\bm{Q}}_{t_k} - q^*) \geq \beta \right) \geq \bbP\left(\sqrt{t_k} (\widehat{\bm{Q}}_{t_k} - \overline{\bm{Q}}_{t_k}) \geq \beta' \right) \geq \text{erfc}\left(\frac{\beta + C}{\gamma(q^*)}\right) > 0.
\end{align}
Consequently, we can use Equation~\eqref{eq:effectiveCLT} to complete the proof-by-contradiction of Theorem~\ref{thm:lastiteratedivergence}.
In particular, we follow an identical series of steps to the proof of Theorem~\ref{thm:warmup} to show that if $(\bm{P}_t, \bm{Q}_t) \to (p^*,q^*) \text{a.s.}$, we need to have
\begin{align*}
\lim_{k \to \infty} \bbP[\bm{P}_{t_k} \geq p^* + \delta] &\geq \lim_{k \to \infty} \bbP\left(\sqrt{t_k} (\widehat{\bm{Q}}_{t_k} - q^*) \geq \beta \right) \geq \text{erfc}\left(\frac{\beta + C}{\gamma(q^*)}\right) > 0,
\end{align*}
where $\beta > 0$ is chosen just as in the proof of Theorem~\ref{thm:warmup}.
This provides the desired contradiction.
Ultimately, we have shown that if the last iterate of player $2$ were assumed to converge, the last iterate of player $1$ could not converge (in fact, even in probability).
This completes our proof-by-contradiction: at least one of the players cannot converge almost surely to NE.
% \Halmos
\endproof
% \end{proof}
%!TEX root = ../main.tex

\subsection{Beyond the mean-based assumption}
\label{sec: beyond_mean_based}

In this section, we show that an \emph{exact} mean-based strategy (Definition~\ref{as:meanbased}) is not required to produce the last-iterate oscillation phenomenon.
% While the mean-based assumption is fairly strong, it is worth noting that mean-based strategies underlie the design of almost all no-regret algorithms in practice.
In particular, we show that equivalents of Theorem~\ref{thm:lastiteratedivergence} hold under two algorithmic variants of mean-based strategies that are ubiquitous in the online learning and games literature.

\subsubsection{Oscillation under optimism}
One of the most common algorithmic variants of exact mean-based strategies incorporates a form of recency bias, colloquially called ``optimism''.
We define the broad class of recency-bias strategies below.
\begin{definition}\label{def:krecencybias}
The class of $\ell$-recency bias strategies is defined as below:
\begin{align*}
f_t(\bm{J^{t-1}}) := f_t(\bm{\widehat{Q}^\ell_{t-1}}) \text{ where } \\
\bm {\widehat{Q}^\ell_t} := \frac{1}{t} \left(\sum_{s=1}^{t} \bm{J_s} + \sum_{j = 1}^{\ell} r_j \bm{J_{t - j + 1}}\right) ,
\end{align*}
and $\{r_j\}_{j=1}^\ell$ are positive integers taking values in $\{1,\ldots, \ell\}$.
We continue to assume that $f_t(\cdot)$ is monotonic in its argument for every $t \geq 1$ (similar to Definition~\ref{as:monotonic}).
\end{definition}

Note that the class of $0$-recency bias strategies is equivalent to the class of mean-based strategies.
Further, $1$-recency bias strategies with $r_1 = 1$ constitute the class of \textit{optimistic} mean-based strategies, since they are using $\sum_{s=1}^t \bm{J_s} + \bm{J_t}$ as the summary statistic.

As mentioned in Section~\ref{sec: intro}, the study of the last iterate of optimism-based strategies (and the related classic extra-gradient method~\cite{korpelevich1976extragradient}) has generated a lot of interest in the optimization literature~\citep{daskalakis2018training,mertikopoulos2018optimistic,liang2019interaction,abernethy2019last,lei2020last}; more-over, these strategies are known to cause faster time-averaged convergence~\citep{rakhlin2013optimization,syrgkanis2015fast}.
Most recently, it was shown that the last iterate of the players' strategies in the setting of telepathic dynamics (that arises when players use each others' mixtures to update their strategies) will converge~\citep{daskalakis2018last}.
In the more realistic realization-based model, the following result shows that the ensuing stochasticity \textit{alone} causes recency-bias-based strategies to diverge.

\begin{theorem}\label{thm:optimism}
Fix any $\ell > 0$.
If both players $1$ and $2$ use any $\ell$-recency-biased, monotonic, optimal-no-regret strategies $\{f_t\}_{t \geq 1}$ and $\{g_t\}_{t \geq 1}$ respectively, then the pair of the mixed strategies of both the players $(\bm{P}_t,\bm{Q}_t)$ cannot converge to $(p^*,q^*)$ almost surely.
\end{theorem}
The proof of Theorem~\ref{thm:optimism} is a relatively simple modification of the proof of Theorem~\ref{thm:lastiteratedivergence}, and is provided below.
\proof{Proof of Theorem~\ref{thm:optimism}.}
We first define notation pertinent to $\ell$-recency-bias strategies.
We denote $\bm{Z^\ell_t} := \sum_{t'=1}^t \bm{J_{t'}} + \sum_{j=1}^{\ell} r_j \bm{J_{t-j + 1}}$, where $\{\bm{J}_t\}_{t \geq 1}$ denotes the sequence of realizations generated by player $2$.
Note that $\bm{\widehat{Q}^\ell_t} = \bm{Z^\ell_t}/t$.
More conveniently, we can also write
\begin{align*}
\bm{Z^\ell_t} := \sum_{t'=1}^t (1 + r'_{t'}) \bm{J_{t'}} ,
\end{align*}
where we designate $r'_{t'} = 0$ for $t' \leq (t - \ell)$, and $r'_{t'} = r_{t - t' + 1}$ thereafter.

We will essentially mimic the proof-by-contradiction approach of Theorem~\ref{thm:lastiteratedivergence}.
We will suppose that $(\bm{P}_t,\bm{Q}_t) \to (p^*,q^*)$ almost surely, and show that if $\bm{Q}_t \to q^*$ almost surely, then $\bm{P}_t$ must oscillate, which provides the desired contradiction.
We state the following claim:
\begin{claim}\label{claim:optimismclt}
Let $C$ be the universal constant (well-defined by Lemma~\ref{lem:timeave} for any pair of optimal no-regret algorithms for players $1$ and $2$) such that $\bm{\overline{Q}_t} \geq q^* - C/\sqrt{t}$ pointwise.
Then, for any deterministic subsequence $\{t_k\}_{k \geq 1}$ and any $\beta > 0$, we have
\begin{align}\label{eq:optimismclt}
\lim_{k \to \infty} \bbP(\sqrt{t_k} (\bm{\widehat{Q}^\ell_{t_k}} - q^*) \geq \beta) \geq \text{erfc}\left(\frac{\beta + C}{\gamma(q^*)}\right) > 0 ,
\end{align}
where $\gamma(\cdot) > 0$ is defined as in the proof of Theorem~\ref{thm:lastiteratedivergence}.
\end{claim}
Claim~\ref{claim:optimismclt} essentially provides a CLT-like-statement for the recency-bias-adjusted random sequence $\{\bm{\widehat{Q}^\ell_{t_k}}\}_{k \geq 1}$, and is proved below.
\proof{Proof of Claim~\ref{claim:optimismclt}.}
% \begin{proof}
% The first step is to prove a CLT-like statement on the differences $\bm{\widehat{Q}^\ell_{t_k}} - \bm{\overline{Q}_{t_k}}$.
By the definition of $\bm{\widehat{Q}^\ell_{t_k}}$, we notice that
\begin{align*}
\bm{\widehat{Q}_{t_k}} \leq \bm{\widehat{Q}^\ell_{t_k}} \leq \bm{\widehat{Q}_{t_k}} + \frac{\ell^2}{t_k} \text{ a.s.},
\end{align*}
which gives us
\begin{align*}
\sqrt{t_k}(\bm{\widehat{Q}_{t_k}} - q^*) \leq \sqrt{t_k}(\bm{\widehat{Q}^\ell_{t_k}} - q^*) \leq \sqrt{t_k}(\bm{\widehat{Q}_{t_k}} - q^*) + \frac{\ell^2}{\sqrt{t_k}} \text{ a.s.}
\end{align*}
Since $\ell$ is assumed to be a constant that does not grow with $t$, we can apply the sandwich theorem to get
\begin{align*}
\sqrt{t_k}(\bm{\widehat{Q}^\ell_{t_k}} - q^*) \to \sqrt{t_k}(\bm{\widehat{Q}_{t_k}} -q^*) \text{ a.s.}
\end{align*}
Substituting Equation~\eqref{eq:martingaleCLTsubsequence} from the proof of Theorem~\ref{thm:lastiteratedivergence} (which used the martingale CLT together with the time-averaged convergence property) into the above yields
\begin{align*}
\lim_{k \to \infty} \bbP(\sqrt{t_k}(\bm{\widehat{Q}^\ell_{t_k}} - q^*) \geq \beta) \geq \text{erfc}\left(\frac{\beta + C}{\gamma(q^*)}\right),
\end{align*}
which completes the proof of the claim.
% \Halmos
\endproof
% \end{proof}
We now use Claim~\ref{claim:optimismclt} to complete the proof of Theorem~\ref{thm:optimism}.
We denote $\{\bm{J'_{t}}\}_{t \geq 1}$ to be a sequence of iid Bernoulli$(q^*)$ random variables.
Using this notation, we can then write, for any $0 \leq s \leq t$,
\begin{align*}
\bm{(Z'')^\ell_{t,s}} &:= \sum_{t'=1}^{t-s} (1 + r'_{t'}) \bm{J'_{t'}} + \sum_{t' = t - s + 1}^t (1 + r'_{t'}) \\
\bm{(Z')^\ell_{t}} &:= \sum_{t'=1}^{t} (1 + r'_{t'}) \bm{J'_{t'}} \\
\bm{(\widehat{Q}'')^\ell_{t,s}} &:= \frac{\bm{(Z'')^\ell_{t,s}}}{t} \\
\bm{(\widehat{Q}')^\ell_{t}} &:= \frac{\bm{(Z')^\ell_{t}}}{t}. \\
\end{align*}
From Proposition~\ref{prop:noregretsensitivity}, we know that there exists a sequence $\{t_k,s_k\}_{k \geq 1}$ such that $0 \leq s_k \leq \alpha (t_k)^{1/2}$ for all $k \geq 1$, and 
\begin{align*}
\bbE\left[f_{t_k}(\bm{(\widehat{Q}'')^\ell_{t_k,s_k}})\right] \geq p^* + 2 \delta \text{ for all } k \geq 1 .
\end{align*}
As in the proof of Theorem~\ref{thm:warmup}, we can use Markov's inequality to get
\begin{align*}
\bbP\left(f_{t_k}(\bm{(\widehat{Q}'')^\ell_{t_k,s_k}}) > p^* + \delta \right) \geq \epsilon_0 ,
\end{align*}
where $\epsilon_0 := \delta/((1 - p^*) - \delta)$.
Note that $0 < \epsilon_0 < 1/2$, as in the proof of Theorem~\ref{thm:lastiteratedivergence}.

Next, in an argument similar to the proof of Claim~\ref{claim:optimismclt}, we can apply the central-limit-theorem to the recency-biased random variable $\bm{(\widehat{Q}')^\ell_t}= \bm{(Z')^\ell_t}/t$.
This will yield
\begin{equation}
\label{eq: prob_markov_tailcut_bdd_opt}
    \bbP \left(f_{t_k} \left(\frac{\bm{({Z}'')^\ell_{t_k, s_k}}}{t_k} \right) \geq p^* + \delta, \bm{(Z'')^\ell_{t_k, s_k}} \leq q^* t_k + \beta \sqrt{t_k} \right) \geq \frac{\epsilon_0}{2}.
\end{equation}
Equation~\eqref{eq: prob_markov_tailcut_bdd_opt} together with the monotonicity assumption (Assumption~\ref{as:monotonic}) yields
\begin{align*}
f_{t_k}\left(q^* + \frac{\beta}{\sqrt{t_k}}\right) \geq p^* + \delta \text{ for all } k \geq 1,
\end{align*}
where the justification is the same as in the proof of Theorem~\ref{thm:warmup}, and $\beta > 0$ is chosen in the same way.
Accordingly, we get
\begin{align*}
\lim_{k \to \infty} \bbP[\bm{P}_{t_k} \geq p^* + \delta] \geq \lim_{k \to \infty} \bbP[\sqrt{t_k}(\bm{\widehat{Q}_{t_k}} - q^*) \geq \beta] \geq \text{erfc} \left(\frac{\beta + C}{\gamma(q^*)}\right),
\end{align*}
where the last inequality follows from Equation~\eqref{eq:optimismclt}.
This completes the proof of Theorem~\ref{thm:optimism}.
% \Halmos
\endproof

It is worth noting that the bounded memory of the recency bias, as well as bounded increments $\{r_j\}_{j=1}^{\ell}$ are critical to the essence of the proof argument.
It is plausible that stronger recency biases that grow with the number of steps $t$ could lead to different last-iterate behavior; however, such stronger recency biases could also break the no-regret property.

\subsubsection{Oscillation under stochastic domination (e.g. adaptive step sizes)}

The assumption of mean-based strategies (Definition~\ref{as:meanbased}) is not satisfied by some of the most popular online learning algorithms that retain the worst-case optimal no-regret guarantee, but obtain faster rates under ``easier data''.
These include variants of Online-Mirror-Descent that adapt to small cumulative loss or variance~\citep{hazan2010extracting}, predictable sequences~\citep{rakhlin2013online,rakhlin2013optimization}, and stochasticity~\citep{cesa2007improved,erven2011adaptive}; while retaining the usual worst-case bounds.
The catch in these algorithms is that the learning rate (or step size) can now be adaptively set, i.e. $\eta_t := \eta_t(\bm{(I)^{t-1}}, \bm{(J)^{t-1}})$.
The adaptive learning rate can be set via the ubiquitous ``doubling-trick"~\citep{cesa2007improved,hazan2010extracting} or in a more continuous manner~\citep{erven2011adaptive,rakhlin2013online,rakhlin2013optimization}.
Either way, none of these algorithms satisfy the mean-based property as in general, the learning rate function $\eta_t(\mathcal{F}_{t-1})$ will depend in some non-linear way on $\{\bm{P}_s,\bm{Q}_s,\bm{I}_s,\bm{J}_s\}_{s=1}^{t-1}$.

In spite of this, we can prove that the last-iterate oscillates even for these algorithms by showing that their iterates \emph{stochastically dominate} those of an algorithm in the Online-Mirror-Descent family, which is mean-based.
More generally, we now show that our main result of last-iterate oscillation would directly extend to \textit{any} algorithm whose iterates stochastically dominate those of a mean-based algorithm.
\begin{corollary}\label{cor:extension}
Consider any optimal-no-regret strategy for player $1$ (or player $2$) $\{f_t\}_{t \geq 1}$ that \textbf{stochastically dominates} some mean-based and monotonic optimal-no-regret strategy $\{f'_t\}_{t \geq 1}$ in the following sense: for every $t \geq 1$ and every history $((I)^{t-1}, (J)^{t-1})$, we have
\begin{align}\label{eq:stochasticdominance}
|f_t((I)^{t-1}, (J)^{t-1})) - p^*| \geq |f'_t(\widehat{Q}_{t-1}) - p^*| .
\end{align}
Further, let player $2$ (or player $1$) follow \textbf{any} optimal-no-regret strategy $\{g_t\}_{t \geq 1}$.
Then, the pair of mixed strategies $(\bm{P_t},\bm{Q_t})$ cannot converge almost surely.
\end{corollary}

Corollary~\ref{cor:extension} automatically implies that the players cannot converge almost surely if even one of them is using any of the aforementioned adaptive variants of Online-Mirror-Descent.
Let us see why this is true for the simplest case of adaptive variants of multiplicative weights/Hedge (which is a special case of the Online-Mirror-Descent family).
We recall the following critical property that is shared by all adaptive variants of Hedge: for some universal constant $C > 0$, we have
\begin{align*}
\eta_t \geq \frac{C}{\sqrt{t}} \text{ pointwise. }
\end{align*}

This property is usually used to prove that such adaptive algorithms retain the $\mathcal{O}(\sqrt{t})$ regret guarantee in the worst case (while admitting faster rates under more favorable conditions).
For a given round $t$, we define $\bm{Z}_{t-1}$ as above, and further define
\begin{align*}
\bm{P}_{t,\mathsf{Hedge}} := f_{t,C}(\widehat{\bm{Q}}_{t-1}) := \frac{\exp\{C \sqrt{t} \widehat{\bm{Q}}_{t-1} \}}{\exp\{C \sqrt{t} \widehat{\bm{Q}}_{t-1} \} + \exp\{C \sqrt{t} (1 - \widehat{\bm{Q}}_{t-1})\}}
\end{align*}
as the \emph{counterfactual iterate} that would have been generated if the algorithm switched to Hedge with learning rate $\eta_t = C/\sqrt{t}$ at round $t$. We note that $|\bm{P}_{t} - 1/2| \geq |\bm{P}_{t,\mathsf{Hedge}} - 1/2|$, implying that the stochastic dominance property in Equation~\eqref{eq:stochasticdominance} is satisfied.

We conclude this section with the proof of Corollary~\ref{cor:extension}, provided below.

%!TEX root = ../main.tex

% \begin{proof}
\proof{Proof of Corollary~\ref{cor:extension}.}
At every round $t$, we can consider the \textit{counterfactual} iterate given by $\bm{P}'_t := f'_t(\bm{\widehat{Q}}_{t-1})$.
Using this, we can follow the same proof-by-contradiction strategy as considered in the proof of Theorem~\ref{thm:lastiteratedivergence}, by observing two facts:
\begin{enumerate}
\item The two essential properties of $\bm{\widehat{Q}}_{t-1}$: that it satisfies the conditions for the martingale CLT, and that it is sufficiently close to the average iterate given by $\bm{\overline{Q}}_{t-1}$, continue to hold here. The former property is a consequence of the assumed convergence that we will contradict, and the latter property is a consequence of both players following optimal-no-regret algorithms.
\item We have $\bm{P}_t \geq \bm{P}'_t$ almost surely.
Since the proof of Theorem~\ref{thm:lastiteratedivergence} uses that $\bm{P}'_t \geq p^* + \delta$ whenever $\bm{\widehat{Q}_{t-1}} \geq q^* + \frac{\beta}{\sqrt{t}}$, it clearly follows that $\bm{P}_t \geq p^* + \delta$ under this event as well.
\end{enumerate}
The proof of Corollary~\ref{cor:extension} trivially follows as a consequence of these facts.
% \Halmos
\endproof
% \end{proof}

We remark that the Online-Mirror-Descent family, while ubiquitous, does not comprise all algorithms used in online prediction in game-theoretic environments.
In particular, algorithms under the umbrella of the ``parameter-free'' online learning paradigm~\citep{chaudhuri2009parameter,orabona2014simultaneous,luo2015achieving,koolen2015second,van2016metagrad} employ a diversity of much more complex updates.
Despite this diversity, all of the updates use some notion of an empirical average across time, where what is being averaged is a functional of the opponent's realized play.
For example, in the case of Squint~\cite{koolen2015second} the update is an empirical average of an instantaneous regret minus a variance term: this is a nonlinear functional of the opponent's realized play.
This nonlinearity implies that algorithms like Squint are not, in the strict sense that we have defined, mean-based.
However, as long as such functionals disambiguate the opponent's pure strategies $\{0,1\}$, we do expect our proof technique to also extend to this class of algorithms, albeit with some additional algebraic work.
(At a high level, this is because the main workhorse for our proof is the CLT, which applies to sums of independent random variables under very mild distributional assumptions.)
More essential to our framework is the requirement of the updates being a function of \emph{some} notion of an empirical average across time.
Removing this empirical-average assumption entirely (i.e. beyond the approximate senses outlined in this subsection) is an important direction for future work.
%!TEX root = ../main.tex

\subsection{Beyond the monotonicity assumption: A conjecture}\label{sec:conjecture}

In Section~\ref{sec: beyond_mean_based}, we showed that our results can be extended beyond exact mean-based strategies, allowing their applicability to a wide range of popular online learning algorithms.
In this section, we ask whether the other principal assumption of \textit{monotonicity} of the mapping from empirical average to mixed strategy at every round (Definition~\ref{as:monotonic}) is also relaxable.
While this assumption of monotonicity will intuitively be satisfied by any learning agent that aims to maximize its utility in a qualitative sense (including expected-utility-theory, but also much broader behavioral models), and can be verified to be satisfied by all known mean-based no-regret algorithms, it is of interest to examine whether it is truly necessary to prove our results.
We conjecture below that last-iterate oscillations will continue to hold for mean-based strategies that are not monotonic.

\begin{conjecture}\label{con:lastiteratedivergence}
If both players $1$ and $2$ use mean-based (not necessarily monotonic) repeated game strategies $\{f_t\}_{t \geq 1}$ and $\{g_t\}_{t \geq 1}$ that are uniformly no-regret and each have a regret rate of $(1/2,c)$, 
then the pair of mixed strategies $(\bm{P_t}, \bm{Q_t})$ does not converge almost surely.
\end{conjecture}
Conjecture~\ref{con:lastiteratedivergence} turns out to be significantly more difficult to prove than the corresponding result with the additional monotonicity assumption, i.e. Theorem~\ref{thm:lastiteratedivergence}.
The assumption of monotonicity on strategy maps (Definition~\ref{as:monotonic}) allowed us to link the event of last-iterate oscillations to an inequality relation on the empirical average $\bm{\widehat{Q}_{t-1}}$; therefore, we could lower bound the probability of a last-iterate oscillation by the cumulative distribution function (CDF) of $\bm{\widehat{Q}_{t-1}}$ and invoke limit laws.
In the absence of monotonicity of strategy maps, it turns out that we need to reason about the \textit{probability mass function} (PMF) of $\bm{\widehat{Q}_{t-1}}$ instead.
This is a mathematically far more difficult object to study under the general scenario where $\bm{Q}_t$ is now a complex functional of the history of $(\bm{I}^{t-1},\bm{J}^{t-1})$.
Importantly, the realizations of player $2$, given by $\{\bm{J_t}\}_{t \geq 1}$ are \textit{not} mutually independent, even conditionally on the mixed strategies $\{\bm{Q_t}\}_{t \geq 1}$.
Moreover, martingale structure by itself is insufficient to obtain adequate control on the PMF of $\bm{\widehat{Q}_{t-1}}$.
% \vidyacomment{This is just way too in-the-weeds now; we need to think about improving the writing here.}

Nevertheless, we can make partial progress on proving Conjecture~\ref{con:lastiteratedivergence}.
First, we note that the proof of the ``warm-up'' Theorem~\ref{thm:warmup} can be modified to show that player $1$ would oscillate in the idealized case where player $2$ has already converged to his NE even when the mean-based strategies are non-monotonic (Proposition~\ref{prop:warmup_ext} in Appendix~\ref{sec:fixedconvergent}).
This is a consequence of the mutual independence of the realizations $\{\bm{J_t}\}_{t \geq 1}$ in this case.
While we know that the realizations $\{\bm{J_t}\}_{t \geq 1}$ cannot be mutually independent in general, they are still likely to be ``minimally stochastic" across rounds in a certain sense: after all, an independent coin is being tossed on every round to generate the realization of player $2$, $\bm{J_t}$, from his mixed strategy on that round, $\bm{Q_t}$.
Thus, it is reasonable to conjecture that the PMF of $\bm{\widehat{Q}_t}$ is sufficiently ``close'' to a sum of independent random variables, which we denote by:
\begin{align*}
\bm{Z}'_t &:= \sum_{s=1}^t \bm{J}'_s \text{ where } \\
\bm{J}'_s &\sim \text{Ber}(\bbE[\bm{Q}_s]), \bm{J'}_s \text{ mutually independent. }
\end{align*}
More precisely, suppose that we could show that the induced distribution on the empirical average of player $2$, i.e. $\bm{\widehat Q_t}$, was similar to the normalized sum of independent random variables, defined above by $\bm{Z_t}'/t$, in the following quantitative sense:
\begin{definition}
The mean-based strategies $\{f_t\}_{t \geq 1}$ and $\{g_t\}_{t \geq 1}$ would satisfy the \textbf{shaky-hands} property if for any $\gamma > 0$, there exists an $\epsilon > 0$ and a $T > 0$ such that
\begin{align}\label{eq: req_measure_ratio_prop}
 \frac{\bbP(t\bm{\hat Q_t} = z)}{\bbP(\bm{Z'_t} = z)} \geq \epsilon \text{  for all } z \in [tq^* - \gamma \sqrt{t}, tq^* + \gamma \sqrt{t}],
\end{align}
for all $t \geq T$.
\end{definition}
This \emph{shaky-hands} property, if true, would posit a minimal amount of stochasticity on the dependent realizations of player $2$.
The following theorem shows that Conjecture~\ref{con:lastiteratedivergence} is true if the shaky-hands property holds.

\begin{theorem}\label{thm:shakyhands}
    If the mean-based strategies $\{f_t\}_{t \geq 1}$ and $\{g_t\}_{t \geq 1}$ satisfy the Shaky-hands property (Equation~\eqref{eq: req_measure_ratio_prop}), and the strategy $\{f_t\}_{t \geq 1}$ of player $1$ has a regret-rate of $(1/2,c)$, then player $1$'s last-iterates diverge from the equilibrium strategy $p^*$ in probability, i.e. there exist positive constants $(\delta,\epsilon_0)$ such that
    \begin{align}
    %\label{eq: liminf_prob_dev_response}
        {\lim \sup}_{t \to \infty} \bbP\left[|\bm{P_{t}} - p^*| \geq \delta \right] \geq \epsilon_0 .
    \end{align}
\end{theorem}
Theorem~\ref{thm:shakyhands} shows, in fact, a stronger statement of lack of convergence \emph{in probability} under the shaky-hands property.
This is a valuable result, as it ensures that proving Conjecture~\ref{con:lastiteratedivergence} reduces to proving that the shaky-hands property (Equation~\eqref{eq: req_measure_ratio_prop}) will be satisfied by an arbitrary mean-based, non-monotonic no-regret algorithm.
The proof of Theorem~\ref{thm:shakyhands} is significantly more technically involved than the CLT-based arguments provided in the previous sections, and is provided in Appendix~\ref{sec:fixedconvergent}.

\section{Simulations}
\label{sec:simulations}

In this section, we provide empirical evidence for last-iterate oscillations under even \emph{suboptimal} no-regret strategies.
We evaluate three popular no-regret strategies:

\begin{enumerate}
\item The standard multiplicative weights update, which is known to lead to last-iterate oscillation even under telepathic dynamics (the deterministic setting)~\citep{bailey2018multiplicative}.
\item The optimistic multiplicative weights update, which converges in the last-iterate in the deterministic setting~\citep{daskalakis2018last}.
\item The online mirror descent algorithm with the log function as regularizer, often called ``log barrier"~\citep{nemirovsky1983problem}.
This regularizer has been successfully used to establish robustness of fast \textit{time-average} convergence guarantees in limited-information feedback settings~\citep{foster2016learning}, and is thus naturally interesting to evaluate.
\end{enumerate}

All of the above algorithms fall under the online-mirror-descent framework, and employ fixed learning rates $\{\eta_t\}_{t \geq 1}$.
The rate of decay of $\eta_t$ with $t$ dictates the no-regret rate in all three cases: for any $r \geq 1/2$, if $\eta_t = 1/t^r$, the no-regret rate is equal to $(r,c)$ for a suitable positive constant $c$.
We will evaluate these algorithms with two learning rate choices: $\eta_t = 1/\sqrt{t}$ (optimal), and $\eta_t = 1/t^{0.7}$ (suboptimal rate\footnote{It is easy to verify that the same suboptimal rate $r = 0.7$ would also result from the slower choice of learning rate decay, $\eta_t = 1/t^{0.3}$. We do not evaluate this choice, as the argument in Proposition~\ref{prop:noregretsensitivity} can be employed to show that it results in even more fluctuation-sensitivity than the case of optimal-no-regret; thus, it provably causes last-iterate oscillation. Intuitively, higher learning rates correspond to less randomness in the mixed strategies as a function of the past, and so more fluctuation-sensitivity.} $r = 0.7$), and study the evolution of player $1$'s iterates until $T := 10^8$ steps.

Furthermore, we will consider the simplest $2 \times 2$ game: the matching pennies game, for which $G(0,0) = G(1,1) = 1$ and $G(0,1) = G(1,0) = 0$ (without loss of generality, player $1$ is the player who wants the actions to match).
Note that the unique mixed-strategy equilibrium of this game is $p^* = q^* = 1/2$.
We will plot the evolution of the mixed strategies of player $1$ with time --- since the matching pennies game is symmetric, player $2$ has similar behavior.

%Evidence for conjecture with MWU, O-MWU and log-barrier MD in matching pennies AND weighted matching pennies games (2 figures)
%Old figures are still in git, just add "_old" to all fig names to see them in the PDF!
\begin{figure}
    \centering
    \begin{subfigure}[b]{0.49\textwidth}
        \includegraphics[width=\textwidth]{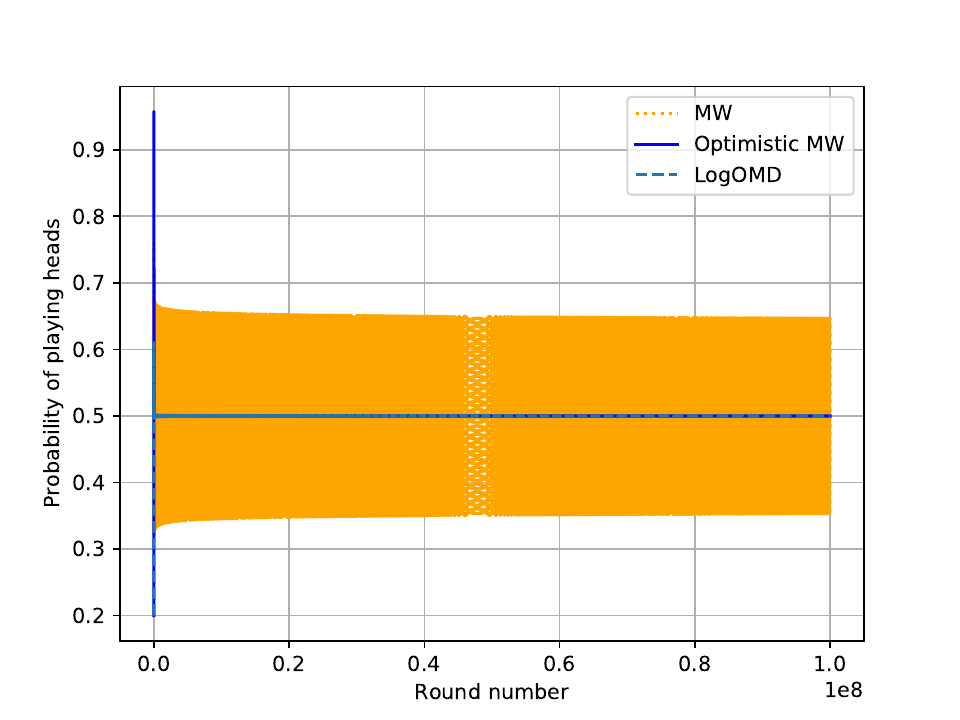}
        \caption{Deterministic case.}
        \label{fig:MPoptimaldeterministic}
    \end{subfigure}%
    ~ %add desired spacing between images, e. g. ~, \quad, \qquad etc.
    \begin{subfigure}[b]{0.49\textwidth}
        \includegraphics[width=\textwidth]{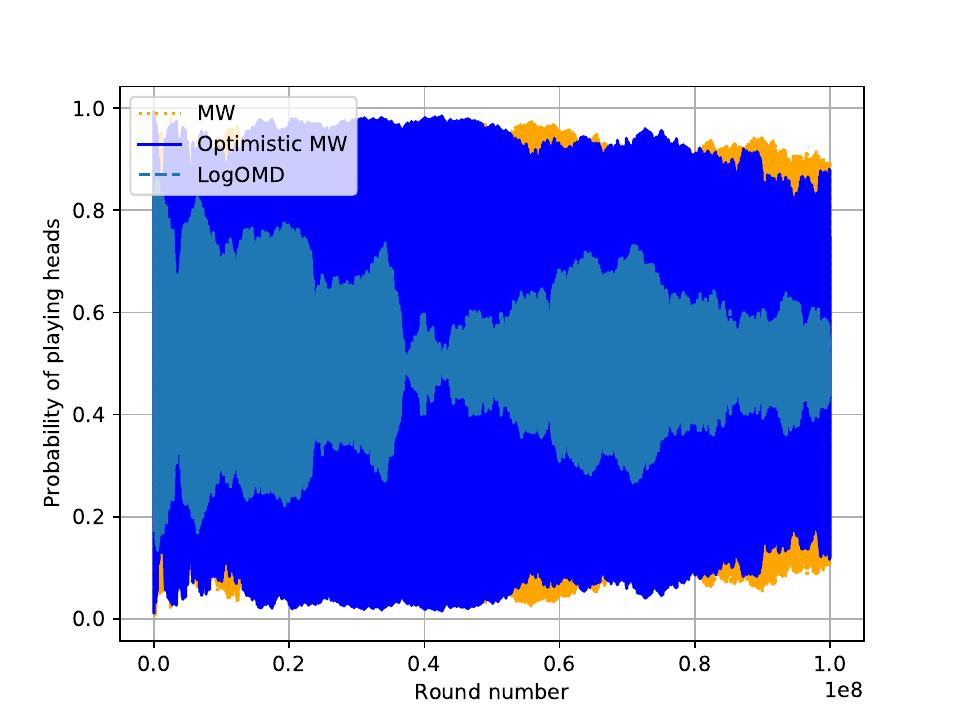}
        \caption{Stochastic case (typical realization).}
        \label{fig:MPoptimalstochastic}
    \end{subfigure}
    \caption{Evolution of the iterates of multiplicative weights in the matching pennies game (for player $1$) when the optimal-no-regret rate $r = 1/2$ is used.}\label{fig:MPoptimal}
\end{figure}

Figure~\ref{fig:MPoptimal} studies the optimal-no-regret case, and shows the striking difference between the evolution of the mixed strategies when the players use opponents' mixtures (the deterministic case) as opposed to their realizations (the stochastic case, studied in this paper).
Notably, we see in Figure~\ref{fig:MPoptimaldeterministic} that while multiplicative weights converges to a limit cycle, optimistic multiplicative weights converges to NE quite quickly.
The third algorithm, log-barrier Online-Mirror-Descent, also oscillates, but the amplitude of the cycles is much smaller than for multiplicative weights, and is in fact not visible in the figure.
This likely reflects the increased entropy of the strategies used in the log-barrier algorithm.
On the other hand, we see in Figure~\ref{fig:MPoptimalstochastic} that all three of these algorithms diverge in the last iterate.
In fact, they are very rarely close to the equilibrium strategy $p^* = 0.5$!
All in all, Figure~\ref{fig:MPoptimalstochastic} empirically corroborates Theorems~\ref{thm:lastiteratedivergence} and~\ref{thm:optimism}, and shows in particular that introducing optimism into no-regret strategies does not fix the issue of last-iterate oscillation.

%Figure still being created, not updated yet
\begin{figure}
    \centering
        \includegraphics[width=0.5\textwidth]{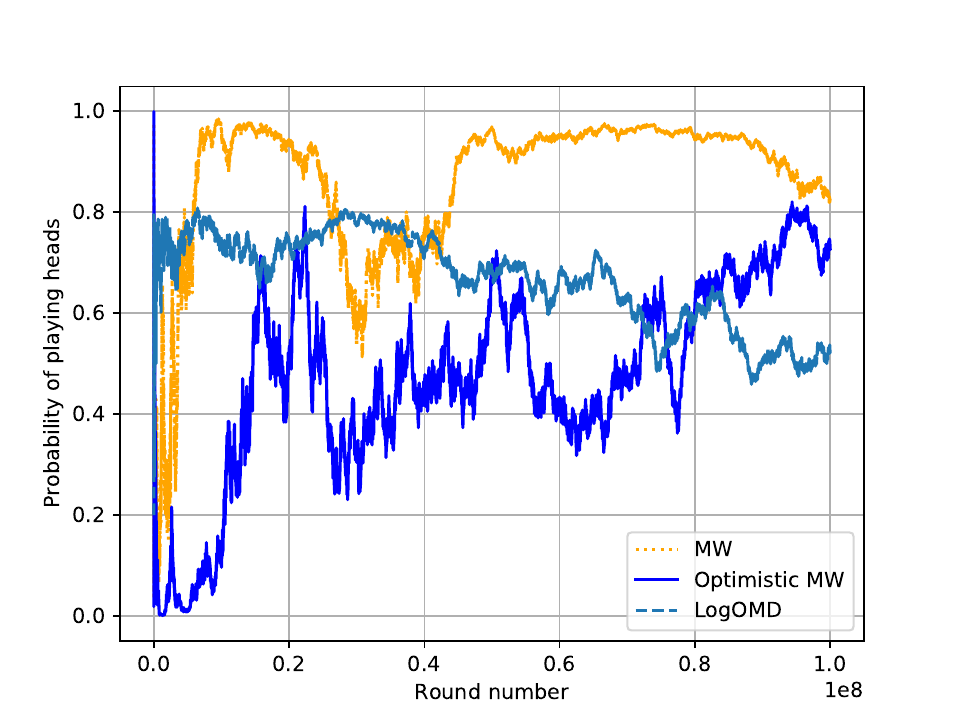}
    \caption{Evolution of the iterates of multiplicative weights in the matching pennies game (for player $1$) when the optimal-no-regret rate $r = 1/2$ is used, and player $2$ is already at NE.}\label{fig:MPoptimalfixedNE}
\end{figure}

It is also worth examining the differential impact on player $1$ as a result of player $2$ using an optimal no-regret strategy, as opposed to player $2$ playing his fixed NE strategy.
In the case of the matching pennies game, the latter case corresponds to player $2$ playing $q^* = 0.5$ at every step.
Figure~\ref{fig:MPoptimalfixedNE} depicts the evolution of the mixed strategies of player $1$ in this latter case.
Comparing the evolution to Figure~\ref{fig:MPoptimalstochastic}, it is evident that the mixed strategies diverge in both cases.
While the ``period'' of limiting cycles, if any, seems to be larger in the fixed-strategy case, the amplitude of oscillation is similar in both cases.
% Thus, the simplifying case that we studied in Theorem~\ref{thm:warmup} successfully identifies at least some of the phenomena underlying last-iterate oscillation.

\begin{figure}
    \centering
    \begin{subfigure}[b]{0.49\textwidth}
        \includegraphics[width=\textwidth]{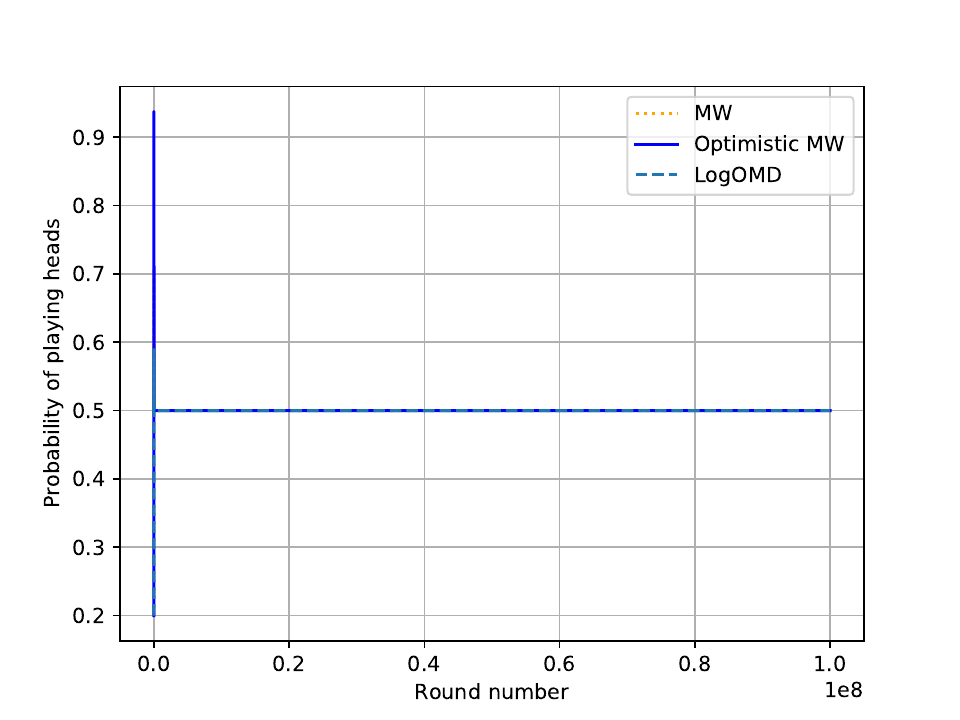}
        \caption{Deterministic case.}
        \label{fig:MPsuboptimaldeterministic}
    \end{subfigure}%
    ~ %add desired spacing between images, e. g. ~, \quad, \qquad etc.
    \begin{subfigure}[b]{0.49\textwidth}
        \includegraphics[width=\textwidth]{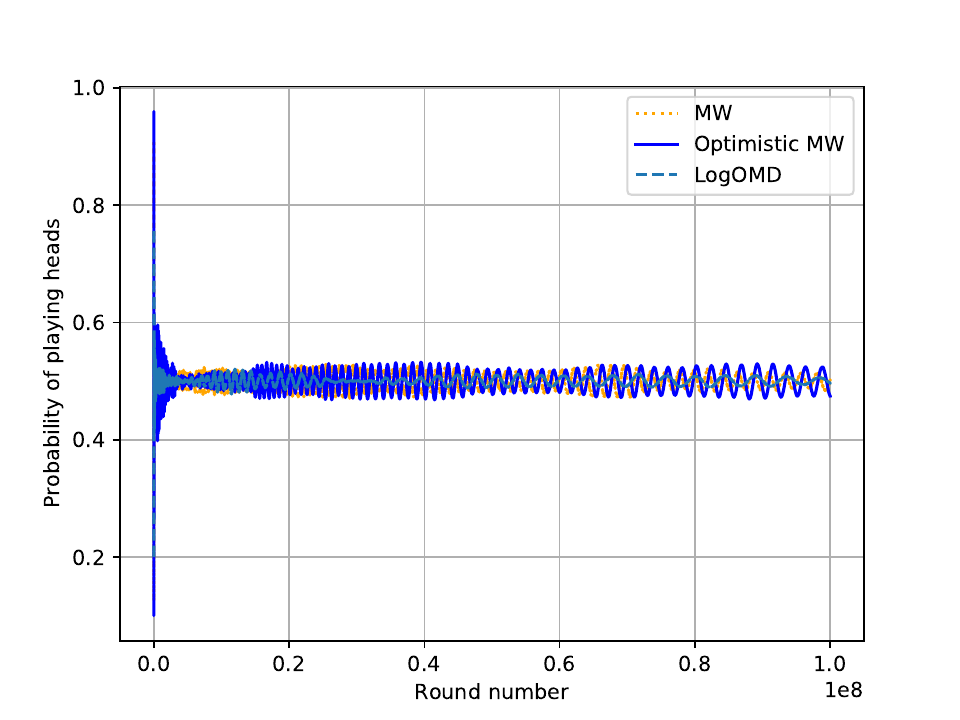}
        \caption{Stochastic case (typical realization).}
        \label{fig:MPsuboptimalstochastic}
    \end{subfigure}
    \caption{Evolution of the iterates of multiplicative weights in the matching pennies game (for player $1$) when the suboptimal-no-regret rate $r = 0.7$ is used.}\label{fig:MPsuboptimal}
\end{figure}

Finally, the techniques in our theory have crucially relied on the optimality of the no-regret algorithms used by both players.
It is naturally interesting to ask whether relaxing the optimality of the no-regret rate could lead to better results in the evolution of the mixed strategies.
Figure~\ref{fig:MPsuboptimal} provides preliminary empirical evidence that this may not be the case.
While Figure~\ref{fig:MPsuboptimaldeterministic} shows the convergence of optimistic multiplicative weights in the telepathic case (as expected), Figure~\ref{fig:MPsuboptimalstochastic} shows that the suboptimal variants of all three algorithms continue to lead to last-iterate oscillation in the stochastic setting (although the amplitude of the oscillations does appear to be sizably reduced).
%!TEX root = ../main.tex

\section{Conclusion and future work}
\label{sec: concl}

In this paper, we have shown partial but compelling evidence for a fundamental tension between the guarantees of no-regret and last-iterate convergence on uncoupled dynamics that use the opponents' realizations alone as feedback.
Perhaps the most important immediate question to address is whether last-iterate oscillations occur for strategies not satisfying monotonicity; in particular, whether Conjecture~\ref{con:lastiteratedivergence} is true.
Additionally, we can ask whether the mean-based nature of the strategies is truly needed for our impossibility result.
While we did show that our results are quite robust to inexact versions of the mean-based assumption (through Theorem~\ref{thm:optimism} and Corollary~\ref{cor:extension}), whether they also hold for strategies that significantly deviate from the class of mean-based strategies is an intriguing question.
% Owing to the mean-based nature of the offline benchmark in regret-minimization, it may even be that all no-regret strategies are in a certain approximate sense mean-based.
% \vidyacomment{This is something I believe, but perhaps it gives too much of future ideas away?}

%The optimal no-regret assumption seems central to our analysis, and suggests that the search for no-regret dynamics with last-iterate convergence should necessarily let go the optimal no-regret criteria.
%This also hints that there might be a fundamental trade-off between no-regret rate and the convergence rate of the strategies.
%We leave this for future work.
Section~\ref{sec:simulations} provided preliminary empirical evidence that relaxing the regret rate may not resolve the issue of last-iterate oscillation, at least for specific algorithms. 
Whether asymptotic convergence becomes possible when suboptimal no-regret algorithms are used is an interesting open question.
Indeed, our techniques cannot be used to show the presence of oscillations in the face of suboptimal no-regret rates, but they can be leveraged to show a lower bound on the \emph{rate} of last-iterate convergence that will imply that \emph{the lower the regret, the slower the rate of convergence.} 
This further suggests that the desiderata of low regret and last-iterate convergence are fundamentally at odds with one another.
% Our techniques break down in the face of suboptimal no-regret rates, so it is interesting to ponder whether last-iterate oscillation happens for general suboptimal no-regret algorithms, or if it is just a property of the particular algorithms that were simulated.

Finally, while last-iterate convergence may seem like a strong requirement in the realization-based repeated game model, we note that there are non-trivial classes of strategies that can be shown to satisfy it.
Of particular interest are the (intractable for large games) \textit{smoothly calibrated} strategies proposed by~\citet{foster2018smooth}.
These strategies constitute randomized responses to deterministic forecasting, and are conceptually quite different from strategies satisfying the no-regret property.
% \vidyacomment{Possibly say something more substantive about this paper here once we've read it fully}
Whether these strategies can be studied more constructively, and from a behavioral game theory standpoint, is an important question for future work.

% \section*{Acknowledgements}
% We thank Venkat Anantharam for useful preliminary discussions.
% VM acknowledges support from a Simons-Berkeley Research Fellowship.
% SP acknowledges the support of NSF grants CNS-1527846, CCF-1618145 and the NSF Science and Technology Center grant CCF-0939370 (Science of Information).
% AS acknowledges support of the ML4Wireless center member companies and NSF grants AST-144078 and ECCS-1343398. 
% Part of this work was done while some of the authors were visiting the Simons Institute for the Theory of Computing.

\appendix
\appendixpage

\section{Connections between non-zero-sum and zero-sum competitive games}\label{sec:competitive}

\begin{figure}
    \centering
    \begin{subfigure}[b]{0.49\textwidth}
        \includegraphics[width=\textwidth]{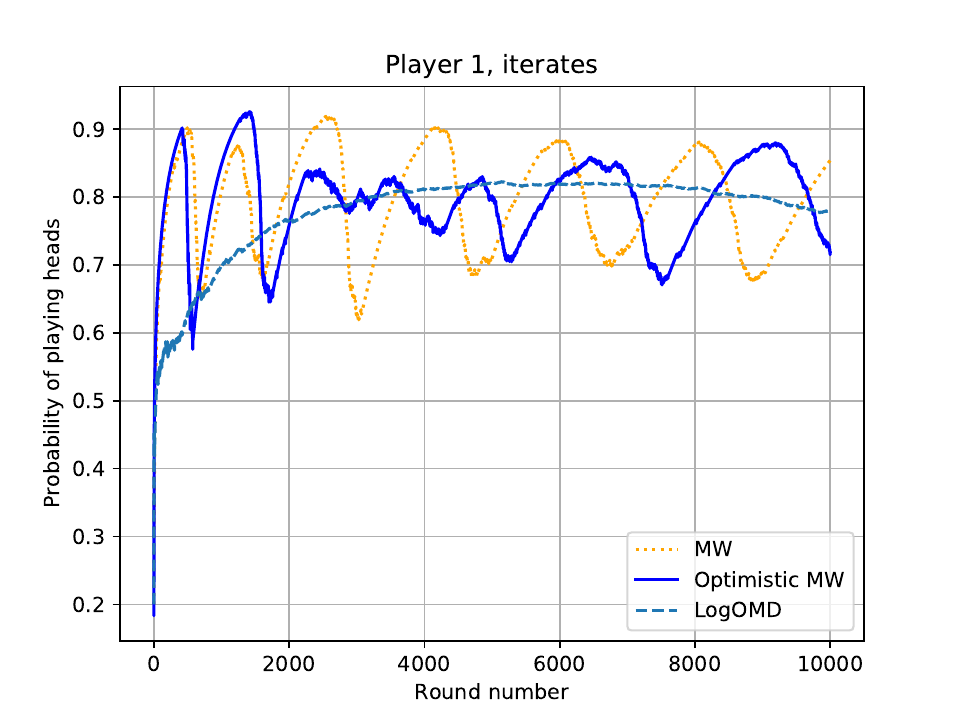}
        \caption{Non-zero-sum game with parameters $(\alpha = 0.111, \beta = 4)$.}
        \label{fig:game1NZS}
    \end{subfigure}%
    ~ %add desired spacing between images, e. g. ~, \quad, \qquad etc.
    \begin{subfigure}[b]{0.49\textwidth}
        \includegraphics[width=\textwidth]{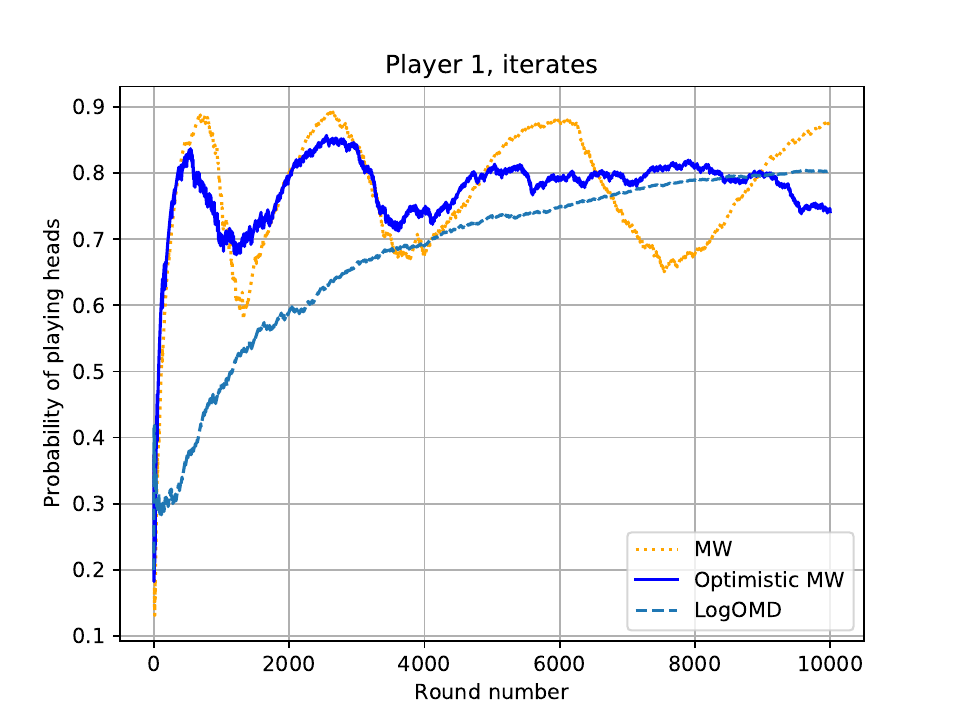}
        \caption{Zero-sum ``equivalent" with parameters $(\alpha = 0.111, \beta = 4)$.}
        \label{fig:game1ZS}
    \end{subfigure}
        \begin{subfigure}[b]{0.49\textwidth}
        \includegraphics[width=\textwidth]{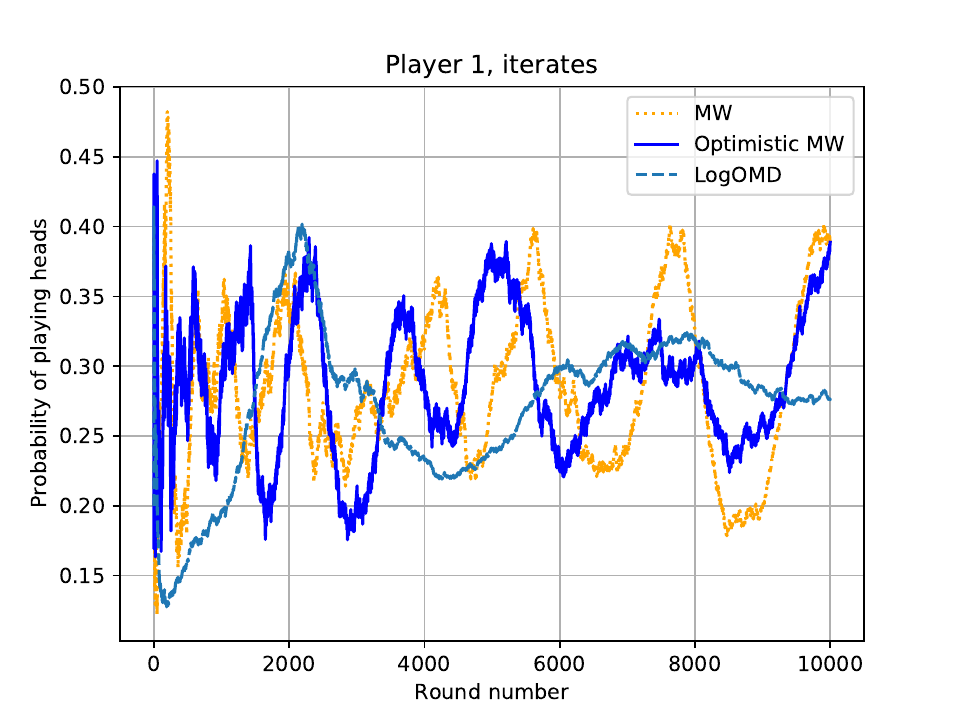}
        \caption{Non-zero-sum game with parameters $(\alpha = 0.25, \beta = 0.429)$.}
        \label{fig:game2NZS}
    \end{subfigure}%
    ~ %add desired spacing between images, e. g. ~, \quad, \qquad etc.
    \begin{subfigure}[b]{0.49\textwidth}
        \includegraphics[width=\textwidth]{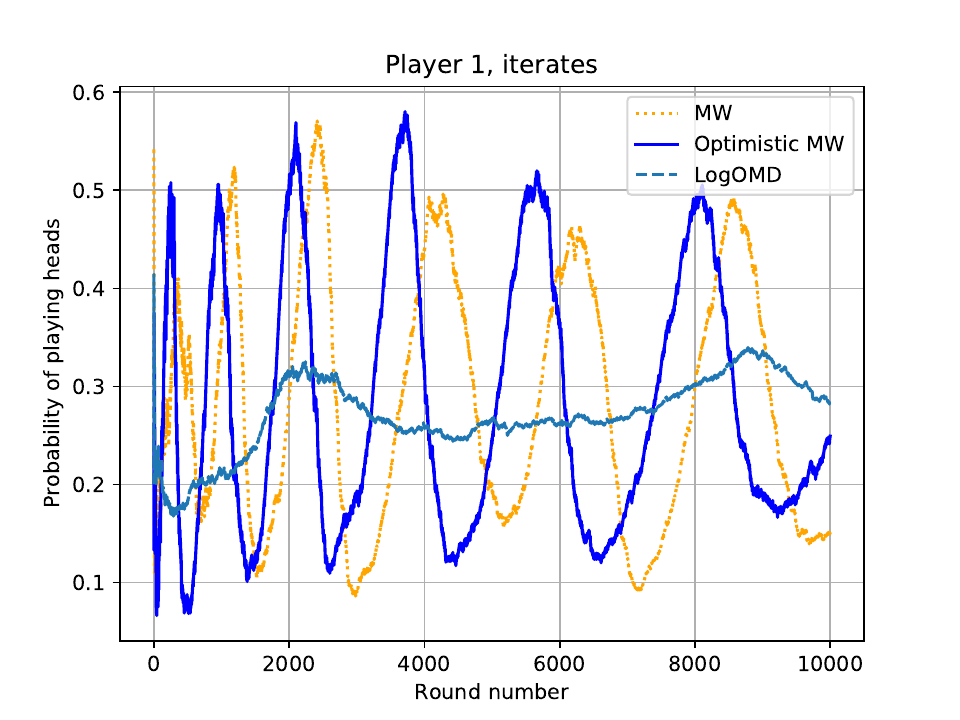}
        \caption{Zero-sum ``equivalent" with parameters $(\alpha = 0.25, \beta = 0.429)$.}
        \label{fig:game2ZS}
    \end{subfigure}
    \caption{Comparison of the iterates of player $1$ using common no-regret algorithms in the non-zero-sum and zero-sum version of the competitive games as specified in Definition~\ref{def:competitivegames2} for two choices of parameters $\alpha,\beta$. While both the non-zero-sum and zero-sum game have identical NE ($p^* = 0.8$ for player $1$ in the first case, and $p^* = 0.3$ for player $1$ in the second case), the iterates as a consequence of no-regret learning exhibit qualitatively different behavior.}\label{fig:competitive}
\end{figure}

In this section, we recap the connection between a non-zero-sum competitive game and a zero-sum competitive game that share the same best-response functions (and thus NE), that was alluded to in Remark~\ref{remark:competitive}.
The following characterization for competitive games was provided in~\citet{phade2019geometry}: 

\begin{definition}\label{def:competitivegames2}
For any choice of parameters $\alpha, \beta > 0$, we define a competitive game with payoff matrix entries given by
\begin{align*}
G(0,0) = -\alpha, G(1,1) &= -1, H(0,0) = \beta, H(1,1) = 1 \text { and } \\
G(0,1) = G(1,0) &= H(0,1) = H(1,0) = 0 .
\end{align*}

The unique Nash equilibrium of this game (which is also a correlated equilibrium) is given by $p^* = \beta/(1 + \beta), q^* = \alpha/(1 + \alpha)$.
Moreover, the corresponding zero-sum game with payoff matrix entries for player $1$ given by 
\begin{align*}
G'(0,0) = \frac{1 - \alpha \beta}{1 + \beta}, G'(0,1) = 1, G'(1,0) = \frac{1 + \alpha}{1 + \beta}, G'(1,1) = 0
\end{align*}
is equivalent to the above competitive game in its best-response function and NE.
Note that the transformation between payoff matrices $(G,H)$ and $(G',-G')$ is non-affine.
\end{definition}

As mentioned in Remark~\ref{remark:competitive}, this equivalence in best-response functions does not translate to an equivalence in actual day-to-day behavior when the players use no-regret algorithms.
See Figure~\ref{fig:competitive} for a depiction of the qualitative discrepancies in behavior for two choices of parameters $(\alpha,\beta)$.
Understanding these discrepancies at a deeper level remains an intriguing direction for future work.

%!TEX root = ../main.tex
\section{Convergence of the empirical average: Proof of Lemma~\ref{lem:timeave}}\label{sec:timeave}
In this section, we provide for completeness the proof of Lemma~\ref{lem:timeave}, which shows almost-sure convergence of the time-average of the mixed strategies of players who deploy no-regret algorithms to the unique mixed NE of a competitive game.
Suppose that both the players employ no-regret algorithms with regret rate $r$. 
Then,~\citet[Proposition~17.9]{roughgarden2016twenty}, shows that the product distribution $\bm{\overline{P}_t} \times \bm{\overline{Q}_t}$ is an approximate \emph{coarse-correlated equilibrium (CCE)}, in the sense that there exists a universal constant $C > 0$ such that
\[
    \bbE_{(\bm{I},\bm{J}) \sim \bm{\overline{P}_t} \times \bm{\overline{Q}_t}}[G(\bm{I},\bm{J})] \leq \bbE_{(\bm{I},\bm{J}) \sim \bm{\overline{P}_t} \times \bm{\overline{Q}_t}}[G(i',\bm{J})] + \frac{C}{T^r} \text{ pointwise}
\]
for any unilateral deviation $i' \in \{0,1\}$ of player $1$,
and similarly,
\[
    \bbE_{(\bm{I},\bm{J}) \sim \bm{\overline{P}_t} \times \bm{\overline{Q}_t}}[H(\bm{I},\bm{J})] \leq \bbE_{(\bm{I},\bm{J}) \sim \bm{\overline{P}_t} \times \bm{\overline{Q}_t}}[H(\bm{I},j')] + \frac{C}{T^r} \text{ pointwise}
\]
for any unilateral deviation $j' \in \{0,1\}$ of player $2$.
Next, we observe that for any competitive game, there is a unique CCE, i.e. there is a unique distribution $\mu$ that satisfies
\[
    \bbE_{(\bm{I},\bm{J}) \sim \mu }[G(\bm{I},\bm{J})] \leq \bbE_{(\bm{I},\bm{J}) \sim \mu }[G(i',\bm{J})],
\]
for any unilateral deviation $i' \in \{0,1\}$ of player $1$, and
\[
    \bbE_{(\bm{I},\bm{J}) \sim \mu}[H(\bm{I},\bm{J})] \leq \bbE_{(\bm{I},\bm{J}) \sim \mu }[H(\bm{I},j')],
\]
for any unilateral deviation $j' \in \{0,1\}$ of player $2$.
In fact, this distribution corresponds to the unique NE of this game, i.e. $\mu = p^* \times q^*$.
As a consequence of this uniqueness, and from the conditions on the payoffs of a competitive game given in Definition~\ref{def: competitive}, we get that the distribution $\mu_t := \bm{\overline{P}_t} \times \bm{\overline{Q}_t}$ is with a distance on the order of $\cal{O}(T^r)$ from $\mu$ in the absolute norm.
In particular, when $r = 1/2$, we get that the time-average of player $2$'s actions converges to $q^*$ at the rate 
\begin{align*}
|\overline{\bm{Q}}_t - q^*| \leq \frac{C}{\sqrt{t}} \text{ pointwise, for all } t \geq 1.
\end{align*}
which is precisely Equation~\eqref{eq:timeave}.
This completes the proof of Lemma~\ref{lem:timeave}.
\qed

%the optimal-no-regret property of player $2$'s strategy to establish the evolution of $\bm{\overline{Q}_t}$ (which is a random variable) as a function of $t$.
While the above reasoning holds for all competitive games and all no-regret algorithms, we note that this lemma was first proved by~\citet{freund1999adaptive} for the special case of zero-sum games with agents playing the multiplicative weights algorithm.
\section{Proof of Theorem~\ref{thm:shakyhands}}\label{sec:fixedconvergent}

In this section, we provide the proof for our partial result (Theorem~\ref{thm:shakyhands}) that shows that last-iterate oscillations can occur even as a consequence of non-monotonic strategies under the conjecture that the shaky-hands condition (Equation~\eqref{eq: req_measure_ratio_prop}) holds.
We first define notation convention specific to this proof.
We designate constants that take a value in $(0,1)$ by $\epsilon$, and strictly positive, finite constants by $C \in (0,\infty)$.
Moreover, we will designate $t_0$ as a lower bound on $t$ above which all our statements apply.
In general, these constants can depend on the parameters of the game, either directly, or just on the equilibrium strategies $(p^*,q^*)$.

For ease of exposition, we will also sub-script these constants by alphabets $\{a,b,\ldots\}$, corresponding to the lemmas in which they appear and are used.
Thus, for example, in the first lemma the constants will be denoted as $\{\epsilon_a,C_a\}$ and the lower bound on $t$ will be denoted as $t_{0,a}$.
While in general we will overload notation within a lemma for our choice of constants, we will be explicit about manipulations when possible.

We begin by showing a version of the ``warm-up'' Theorem~\ref{thm:warmup} that holds for non-monotonic strategies.
\begin{proposition}\label{prop:warmup_ext}
    Let player $2$'s strategy $\{\bm{J_t}\}_{t \geq 1}$ be an i.i.d. sequence of Bernoulli($q^*$) random variables.
    Then, any mean-based repeated game strategy $\{f_t\}_{t \geq 1}$ that has a regret rate of $(1/2,c)$ causes player $1$'s last iterate to diverge, i.e. there exist positive constants $(\delta,\epsilon)$ such that
    \begin{align}
    \label{eq: liminf_prob_dev_response}
        {\lim \sup}_{t \to \infty} \bbP\left[|\bm{P_{t}} - p^*| \geq \delta \right] \geq \epsilon .
    \end{align}
\end{proposition}
% \begin{proof}
\proof{Proof of Proposition~\ref{prop:warmup_ext}.}
The proof is identical to the proof of Theorem~\ref{thm:warmup} until the step
\begin{align*}
    \bbP \left(f_{t_k} \left(\frac{\bm{{Z}''_{t_k, s_k}}}{t_k} \right) \geq p^* + \delta, \bm{Z''_{t_k, s_k}} \leq q^* \cdot t_k + \beta \sqrt{t_k} \right) \geq \frac{\epsilon_0}{2}.
\end{align*}
Since our strategy $\{f_t\}_{t \geq 1}$ is no longer guaranteed to be monotonic, we can no longer turn the above equation into a deterministic statement on which we can apply the central-limit-theorem.
We now use a more specialized argument that controls the ratio of probability mass functions.
Recall that we defined the opponent sequence $\{\bm{J'_t}\}_{t \geq 1}$ to be an iid sequence of Bernoulli($q^*$) random variables.
Note that $\bm{Z'_t} \overset{d}{=} t \bm{\widehat{Q}_{t}} $, and by definition $\bm{Z''_{t,s}} \geq s$ point-wise.
We denote $\beta_0 := \beta/q^*$.
Note that $q^* t + \beta \sqrt{t} = q^* (t + \beta_0\sqrt{t})$ for any $t$.
Now, we show that
    \begin{equation}
    \label{eq: prob_ratio_lower_bdd}
        \min_{s \leq z \leq q^*(t + \beta_0 \sqrt{t}) }  \; \frac{\bbP(\bm{Z'_t} = z)}{\bbP(\bm{Z''_{t,s}} = z)} \geq (1 + \beta_0)^{-\alpha},
    \end{equation}
for all $0 < s \leq \alpha \sqrt{t}$, $t \geq 1$.
Indeed, we have,
    \begin{align*}
        \frac{\bbP(\bm{Z'_t} = z)}{\bbP(\bm{Z''_{t,s}} = z)} 
        &= \frac{\binom{t}{z}(q^*)^z (1 - q^*)^{t - z}}{\binom{t-s}{z-s}(q^*)^{z-s} (1 - q^*)^{t - z}}
        = \frac{t}{z} \cdot \frac{t-1}{z-1} \cdots \frac{t-s+1}{z-s+1} \cdot (q^*)^s\\
        &\geq \left(\frac{q^*t}{z}\right)^s
        \geq \left(\frac{t}{t + \beta_0 \sqrt{t}}\right)^{\alpha \sqrt{t}}\\
        &\geq (1 + \beta_0)^{-\alpha} > 0,
    \end{align*}
where the first inequality follows from $z \leq t$ and therefore $\frac{t - \ell}{z - \ell}$ is increasing in $\ell$, and the last inequality follows from the fact that
    \[
        \left(\frac{t + \beta_0 \sqrt{t}}{t}\right)^{\alpha \sqrt{t}} = \left(1 + \frac{\beta_0}{\sqrt{t}}\right)^{\alpha \sqrt{t}} \leq (1 + \beta_0)^{\alpha}.
    \]
    (Note that the function $(1 + \beta_0/x)^{\alpha x}$ is decreasing in $x$ for $x \geq 1$.)

We are now ready to complete our proof via a simple ``change-of-measure" argument and the above lower bound on the ratio of the probability mass functions.
From Equations~\eqref{eq: prob_markov_tailcut_bdd} and \eqref{eq: prob_ratio_lower_bdd} and the law of total probability, we get
\begin{align*}
    &\bbP \left(f_{t_k} \left(\frac{\bm{{Z}'_{t_k}}}{t_k} \right) \geq p^* + \delta, \bm{Z'_{t_k}} \leq q^* \cdot t_k + \beta \sqrt{t_k} \right) \\
    &\geq \sum_{z=s_k}^{q^* \cdot t_k + \beta \sqrt{t_k}} \bbP \left(\bm{Z'_{t_k}} = z\right) \cdot \mathbb{I}\left[f_{t_k}\left(\frac{z}{t_{k}}\right) \geq p^* + \delta\right]\\
    &\geq (1 + \beta_0)^{-\alpha} \cdot \sum_{z=s_k}^{q^* \cdot t_k + \beta\sqrt{t_k}} \bbP \left(\bm{Z''_{t_k,s_k}} = z\right) \cdot \mathbb{I}\left[f_{t_k}\left(\frac{z}{t_{k}}\right) \geq p^* + \delta\right] \\
    &= (1 + \beta_0)^{-\alpha} \cdot \bbP \left(f_{t_k} \left(\frac{\bm{{Z''}_{t_k}}}{t_k} \right) \geq p^* + \delta, \bm{Z''_{t_k}} \leq q^* \cdot t_k + \beta \sqrt{t_k} \right) \\
    &\geq \frac{\epsilon_0}{2} (1+\beta_0)^{-\alpha},
\end{align*}
and hence
\[
    \bbP \left(f_{t_k} \left(\frac{\bm{{Z}'_{t_k}}}{t_k} \right) \geq p^* + \delta\right) \geq \frac{\epsilon_0}{2} (1+\beta_0)^{-\alpha}, 
\]
for $k \geq k_1$.
Since $\bm{P_{t_k}} \overset{d}{=} f_{t_k}(\bm{\widehat{Q}_{t_k}})$, taking $\epsilon := (\epsilon_0/2) (1+\beta_0)^{-\alpha}$, we get
 \begin{align*}
    \bbP\left[\bm{P_{t_k}} \geq p^* + \delta \right] \geq \epsilon,
    \end{align*}
    for all $k \geq k_1$.
    This implies Equation~\eqref{eq: liminf_prob_dev_response} and completes the proof of the proposition.
% \Halmos
\endproof
% \end{proof}
Now, to extend the ideas from the proof of Proposition~\ref{prop:warmup_ext}, we need to show that the probability mass function of the random variable $\bm{Z_t} := \sum_{s=1}^t \bm{J_s}$ is sufficiently similar to the probability mass function of the sum of independent Bernoulli$(q^*)$ random variables, which we denoted above by $\bm{Z'_t}$.
In other words, it suffices to show that 
\begin{align}\label{eq:pmfratio}
 \frac{\bbP(\bm{{Z}_t} = z)}{\bbP(\bm{Z'_t} = z)} \geq \epsilon \text{ for all } z \in [tq^* - \beta\sqrt{t}, tq^* + \beta\sqrt{t}]
\end{align}
for some universal positive constant $\epsilon > 0$.
We will do this by defining two sets of intermediate random variables:
\begin{enumerate}
\item The random variable $\bm{Y_t} := \sum_{s=1}^t \bm{J''_s}$, where $\bm{J''_t} \sim \text{Ber}(\bbE[\bm{Q_t}])$ and $\{\bm{J''_t}\}_{t \geq 1}$ are mutually independent.
\item The random variable $\bm{Y'_t} := \text{Binomial}(t,\bbE[\bm{\overline{Q}_t}])$.
\end{enumerate}
We will show that the ratios of the pmfs between $(\bm{Z_t}, \bm{Y_t})$, $(\bm{Y_t},\bm{Y'_t})$, and $(\bm{Y'_t}, \bm{Z'_t})$ are lower bounded under the shaky-hands conjecture, i.e. assuming that Equation~\eqref{eq: req_measure_ratio_prop} holds.
First, we note that Equation~\eqref{eq: req_measure_ratio_prop} directly yields
\begin{align*}
 \frac{\bbP(\bm{Z_t} = z)}{\bbP(\bm{Y_t} = z)} \geq \epsilon \text{  for all } z \in [tq^* - \beta \sqrt{t}, tq^* + \beta \sqrt{t}].
\end{align*}
It remains to show that the pmf of $\bm{Y_t}$ is sufficiently similar to the pmf of $\bm{Z'_t} \sim \text{Binomial}(t,q^*)$.
Henceforth, we denote $q_t := \bbE[\bm{Q_t}]$ and $\overline{q}_t := \bbE[\bm{\overline{Q}_t}]$ as shorthand.
We will use a similar proof-by-contradiction approach to the proof of Theorem~\ref{thm:lastiteratedivergence} to establish the required pmf control.
In other words, we will suppose that $\bm{Q_t} \to q^*$ (in fact in probability), and show that we cannot have $\bm{P_t} \to p^*$ in probability.
Note that $\bm{Q_t} \to q^*$ (in probability) obviously requires $q_t \to q^*$.
Moreover, it follows from the property of time-averaged convergence and the definition of a competitive game (Appendix~\ref{sec:timeave}) that $|\overline{q}_t - q^*| \leq \frac{C}{\sqrt{t}}$ for all $t \geq 1$.
In summary, the sequence $\{\bbE[\bm{Q_t}]\}_{t \geq 1}$ belongs to the following class of \textit{fixed-convergent} strategies defined with respect to $q^*$:
\begin{subequations}
\begin{align}
|q_t - q^*| &\leq \delta/2 \text{ for all } t \geq t_0 \label{eq:lastiterateoblivious} \\
|\overline{q}_t - q^*| &\leq \frac{C}{\sqrt{t}} \text{ for all } t \geq 1 \text{, where } \label{eq:timeaverageoblivious} \\
\overline{q}_t &:= \frac{1}{t} \sum_{s=1}^t q_s \nonumber.
\end{align}
\end{subequations}
We also denote the set of such fixed-convergent strategies by $\mathcal{Q}_{\delta,t_0,C}$ and denote their truncation to step $t$ by $\mathcal{Q}_{\delta,t_0,C}(t)$.
% Recall that we earlier defined random variables $\bm{Z_t} \sim \text{Binomial}(t, q^*)$, for $t \geq 1$.
% %, and $\bm{Z''_{t, s}} \sim \bm{Z_{t-s}} + s$.
% Our main proof strategy is to show that the ensuing random variables from any fixed-convergent sub-sequence highly resemble the ensuing random variables from a sequence that is always at equilibrium.
% We consider the sequence $\{\bm{\widetilde{J}_t}\}_{t \geq 1}$ generated by independent random draws from the fixed-convergent strategy of player $2$, i.e. we have $\bm{\widetilde{J}_t} \sim \text{Bernoulli}(q_t)$ for all $t \geq 1$.
% Define a new random variable, $\bm{\widetilde{Z}_t} := \sum_{s=1}^t \bm{\widetilde{J}_s}$.

First, we show that the pmf of $\bm{Y_t}$ is very close to the pmf of  $\bm{Y'_t} \sim \text{Binomial}(t, \bar q_t)$ for any fixed-convergent sequence $\{q_t\}_{t \geq 1}$ satisfying Equations~\eqref{eq:lastiterateoblivious} and~\eqref{eq:timeaverageoblivious} (note that $\bar q_t = \sum_{s = 1}^t q_s$.)
This is encapsulated in the following lemma.

\begin{lemma}\label{lem:changeofmeasure}
Consider any fixed-convergent sequence $\{q_t\}_{t \geq 1}$, i.e. such that Equations~\eqref{eq:lastiterateoblivious} and~\eqref{eq:timeaverageoblivious} both hold.
Let $\bm{Y_t} = \sum_{s=1}^t \bm{J''_s}$ where $\bm{J''_s} \sim \text{Ber}(q_s)$ and $\{\bm{J''_t}\}_{t \geq 1}$ are mutually independent, and let $\bm{Y'_t} \sim \text{Binomial}(t, \bar q_t)$.
Then, there exists positive constant $\epsilon_b$ and integer $t_{0,b}$ such that for all $t \geq t_{0,b}$, we have
\begin{align*}
 \frac{\bbP(\bm{Y_t} = z)}{\bbP(\bm{Y'_t} = z)} \geq \epsilon_b \text{  for all } z \in [tq^* - \beta \sqrt{t}, tq^* + \beta \sqrt{t}] .
\end{align*}

\end{lemma}

We next show that the pmfs of $\bm{Y'_t} \sim \text{Binomial}(t,\overline{q_t})$ and $\bm{Z'_t} \sim \text{Binomial}(t,q^*)$ are sufficiently close.
\begin{lemma}
\label{lem: ration_YtoZ}
    There exists a positive constant $\epsilon_d > 0$ and a sufficiently large $t_{0,d}$ such that for all $t > t_{0,d}$, we have
    \[
        \frac{\bbP(\bm{Y'_t} = z)}{\bbP(\bm{Z'_t} = z)} \geq \epsilon_d, \text{ for all } z \in [q^*t - \beta\sqrt{t}, q^*t + \beta\sqrt{T}].
    \]
\end{lemma}

Notice that from Lemma~\ref{lem:changeofmeasure}, Lemma~\ref{lem: ration_YtoZ} and Equation~\eqref{eq: req_measure_ratio_prop}, we get
\begin{align*}
 \frac{\bbP(\bm{Z_t} = z)}{\bbP(\bm{Z'_t} = z)} \geq \epsilon \text{ for all } z \in [tq^* - \beta\sqrt{T}, tq^* + \beta\sqrt{T}] .
\end{align*}
for a constant $\epsilon = \epsilon_b \epsilon_d > 0$.
In the following two subsections, we prove these two lemmas.
This then completes the proof of Theorem~\ref{thm:shakyhands}.
% \Halmos
% \qed

It remains to prove Lemmas~\ref{lem:changeofmeasure} and~\ref{lem: ration_YtoZ}, which are significantly more technically involved than the martingale CLT that could be applied under the monotonicity assumption.
We denote the pdf of the normal distribution $\mathcal{N}(\mu, \sigma^2)$ by $p(\cdot;\mu,\sigma^2)$.
To prove these lemmas, we use the following theorem of De-Moivre and Laplace in several lemmas that follow.
\begin{theorem}[de-Moivre and Laplace, statement from~\citet{feller1957introduction}]
\label{thm:demoivrelaplace}
Let $\bm{X} \sim \text{Binomial}(t,q)$ for any $0 < q < 1$, and consider any sequence $\{k_t\}_{t \geq 1}$ such that $k_t^3/t^2 \to 0$ as $t \to \infty$.
Then, for every $0 < \epsilon < 1$, there exists a $t_0$ sufficiently large such that for all $t > t_0$, we have
    \[
        1 - \epsilon < \frac{\bbP(\bm{X} = z)}{p(z;tq, tq(1-q))} \leq 1 + \epsilon, \text{ for all integers } qt - k_t \leq  z \leq qt + k_t.
    \]
\end{theorem}
Note that Theorem~\ref{thm:demoivrelaplace} is a much sharper form of asymptotic normality than the typically stated Central Limit Theorem, as it obtains direct control on the probability mass function itself.
% \begin{proof}[Proof of Theorem~\ref{thm:shakyhands}]

\subsection*{Proof of Lemma~\ref{lem:changeofmeasure}}

%\begin{proof}[Proof of Lemma~\ref{lem:changeofmeasure}]
We consider the constant step index $t_{0,b} := \max\{t_0, 4C^2/\delta^2\}$, and note that by the triangle inequality and the assumed convergence (that we are going to contradict), we have
\begin{align*}
|q_t - \overline{q_t}| &\leq |q_t - q^*| + |q^* - \overline{q_t}| \\
&\leq \frac{\delta}{2} + \frac{C}{\sqrt{t}} \\
&\leq \frac{\delta}{2} + \frac{\delta}{2} = \delta ,
\end{align*}
where the last inequality holds for all $t \geq t_{0,b}$.
Thus, we have $|q_t - \overline{q_t}| \leq \delta$, which is useful and required for comparing the pmfs of the random variables $\bm{Y_t}$ and $\bm{Y'_t}$.

Consider a fixed $t \geq t_{0,b}$.
In general, relating the probability mass function of $\bm{Y_t}$, which is the Poisson binomial random variable, directly to the binomial distribution is challenging.
The following technical lemma characterizes the sequence $\{q_s\}_{s=1}^t$ that \textit{minimizes} the probability mass function $\bbP(\bm{Y_t} = z)$ for a fixed choice of $z$.
This minimizing sequence takes values $q_s \in \{\overline q_t - \delta, \overline q_t, \overline q_t + \delta\}$, which turns out to be a much simpler form to analyze.

% \begin{sublemma}
\begin{lemma}
\label{lem:extremizing}
Consider any step index $t \geq 1$.
Then, for every $z \in \{1,\ldots, t\}$, there exists an even integer $0 \leq n_t(z)\leq t$ such that
\begin{align*}
\bbP(\bm{Y_t} = z) \geq \bbP(\bm{\widetilde{Y}_t} = z) ,
\end{align*}
where $\bm{\widetilde{Y}_t} = \text{Binomial}\left(\frac{n_t(z)}{2}, q^* + \delta\right) + \text{Binomial}\left(\frac{n_t(z)}{2}, q^* - \delta\right) + \text{Binomial}\left(t - n_t(z), q^*\right)$.
(Here, the three random variables are independent.)
\end{lemma}
% \end{sublemma}

\proof{Proof of Lemma~\ref{lem:extremizing}.}
%Consider a fixed $T \geq T_2'$.
Let $\eta_s := q_s - \overline q_t$, for $1 \leq s \leq t$, denote the deviation of $q_s$ from the average at time $t$, $\overline{q_t} $.
Thus we have $\sum_{s = 1}^t \eta_s = 0$, and $\eta_s \in [-\delta, \delta]$ for all $s \in \{1,\ldots, t\}$.

Let $e^t := \{e_1, e_2, \dots, e_t\}$ where $e_s \in \{-1, +1\}$ for all $1 \leq s \leq t$ represent the unique encoding of the output sequence $J^t \in \{0,1\}^t$.
Let $|e^t| := |\{e_s = +1 : 1 \leq s \leq t\}|$ denote the number of positive ones in the vector $e^t$.
Now, we consider $1 \leq z \leq t$.
We have
\begin{align*}
    \bbP(\bm{Y_t} = z) &= \sum_{|e^t| = z} \prod_{s = 1}^t q_s \1\{e_s = +1\} + (1 - q_s)\1 \{e_s = -1\})\\
    &= \sum_{|e^t| = z} \prod_{s = 1}^t (\overline{q_t} + \eta_s)\1\{e_s = +1\} + (1 - \overline{q_t} - \eta_s)\1 \{e_s = -1\}).\\
\end{align*}
On the other hand, we have
\begin{align*}
    \bbP(\bm{\widetilde{Y}_t} = z) &= \sum_{|e^t| = z} \prod_{s = 1}^t (\overline{q_t})\1\{e_s = +1\} + (1 - \overline{q_t})\1 \{e_s = -1\})\\
    &= \binom{t}{z} (\overline{q_t})^z (1 - \overline{q_t})^{(t - z)}.
\end{align*}
Thus, we get
\begin{align*}
    \frac{\bbP(\bm{Y_t} = z)}{ \bbP(\bm{\widetilde{Y}_t} = z)} 
    &= \binom{t}{z}^{-1} \sum_{|e^t| = z} \prod_{s = 1}^t \frac{\overline{q_t} + \eta_s}{\overline{q_t}}\1\{e_s = +1\} + \frac{1 - \overline{q_t} - \eta_s}{1 - \overline{q_t}}\1 \{e_s = -1\})\\
    &= \binom{t}{z}^{-1} \sum_{|e^t| = z} \prod_{s = 1}^t \left(1 + \frac{\eta_s}{\overline{q_t}}\right)\1\{e_s = +1\} + \left(1 - \frac{\eta_s}{1 - \overline{q_t}}\right)\1 \{e_s = -1\}).
\end{align*}
Let $\widehat e_s = \frac{+1}{\overline{q_t}}$ if $e_s = +1$ and $\widehat e_s = \frac{-1}{1 - \overline{q_t}}$ if $e_s = -1$.
Let $\widehat e^t = \{\widehat e_1, \dots, \widehat e_t\}$ and let $|\widehat e^t| := |\{\widehat e_s = +1/\overline{q_t} : 1 \leq s \leq t\}|$.
Then, we get
\begin{equation}
    \frac{\bbP(\bm{Y_t} = z)}{ \bbP(\bm{\widetilde{Y}_t} = z)} = \binom{t}{z}^{-1} \sum_{|\widehat e^t| = z} \prod_{s = 1}^t (1 + \widehat e_s \eta_s) .
\end{equation}

We will now try to lower bound the ratio ${\bbP(\bm{Y_t} = z)}/{ \bbP(\bm{\widetilde Y_t} = z)}$ over $\eta^t := \{\eta_1, \dots, \eta_t\}$ such that $\eta_s \in [-\delta, \delta]$ for all $1 \leq s \leq t$ and $\sum_{s = 1}^t \eta_s = 0$. Let $F$ denote the set of all such vectors $\eta^t$.
Let 
\[
    P(\eta^t) = \sum_{|\widehat e^t| = z} \prod_{s = 1}^t (1 + \widehat e_s \eta_s), 
\]
for $\eta \in F$, and let
\[
   \widetilde \eta^t \in \arg \min_{\eta^t \in F} P(\eta).
\]
Note that $P(\eta)$ is a multinomial in $\eta_1, \dots, \eta_t$.
We now show that $\widetilde \eta^t$ satisfies: $\widetilde \eta_s \in \{-\delta, 0, \delta\}$, for all $1 \leq s \leq t$.
First note that if $\widetilde \eta_s \in \{-\delta, \delta \}$ for all $1 \leq s \leq t$ then we are done.
If this does not hold, then without loss of generality let $\widetilde \eta_t \in (-\delta, \delta)$.
Since $\sum_{s = 1}^t \widetilde \eta_s = 0$, let us substitute $\widetilde \eta_t = -\sum_{s = 1}^{t-1} \widetilde \eta_s$.
We now argue that $\widetilde \eta_1 \in \{-\delta, \delta \}$.
We have
\begin{align*}
    &\sum_{|\widehat e^t| = z} \prod_{s = 1}^t (1 + \widehat e_s \widetilde \eta_s) = \sum_{|\widehat e^t| = z} (1 + \widehat e_1 \widetilde \eta_1) (1 - \widehat e_t ( \widetilde \eta_1 + \dots + \widetilde \eta_{t-1}) \prod_{s = 2}^{t-1} (1 + \widehat e_s  \widetilde \eta_s) \\
    &= \sum_{|\widehat e^t| = z} (1 + \widehat e_1 \widetilde \eta_1 - \widehat e_t  \widetilde \eta_1 - \widehat e_t( \widetilde \eta_2 + \dots + \widetilde \eta_{t-1}) - \widehat e_1 \widehat e_t \widetilde \eta_1^2 - \widehat e_1 \widehat e_2 ( \widetilde \eta_2 + \dots +  \widetilde \eta)) H(\widehat e_2, \dots, \widehat e_{t-1}),
\end{align*}
where
\[
    H(\widehat e_2, \dots, \widehat e_{t-1}) = \prod_{s = 2}^{t-1} (1 + \widehat e_s \widetilde \eta_s). 
\]
Note that the above is a quadratic expression in $\widetilde \eta_1$. 
We now observe that the coefficient of $\widetilde \eta_1$ in this expression is zero.
Indeed, the coefficient of $\widetilde \eta_1$ is given by
\begin{align*}
    \sum_{|\widehat e| = z} H(\widehat e_2, \dots, \widehat e_{t-1}) (\widehat e_1 - \widehat e_t) = 0,
\end{align*}
because of the symmetry in $\widehat e_1$ and $\widehat e_t$ in the above expression.
A quadratic of the form $ax^2 + b$ attains its minimum on an interval $[l,h]$ either at $x = l, h$ or $x = 0$.

This establishes that $\widetilde \eta_1 \in \{-\delta, 0, \delta\}$, and indeed the same argument works for all $t \in \{1,\ldots, (T - 1)\}$.
Moreover, we get $\widetilde \eta_t \in \{-\delta, 0, \delta\}$, as these are the only choices that can allow $\sum_{s=1}^t \widetilde \eta_s = 0$.

Thus, we have established that $\widetilde \eta_s \in \{-\delta, 0, \delta\}$ for all $s \in \{1,\ldots, t\}$.
 
Thus, there must be exact $n_t(z)/2$ values of $s$ corresponding to $\widetilde \eta_s = \delta$, $n_t(z)/2$ values of $s$ corresponding to $\widetilde \eta_s = -\delta$, and $(t - n_t(z))$ values of $s$ corresponding to $\widetilde \eta_s = 0$.
Thus, we have shown that
\begin{align*}
\bbP(\bm{Y_t} = z) \geq \bbP(\bm{\widetilde{Y}_t} = z),
\end{align*}
which completes the proof of the lemma.
% \Halmos
\endproof
% \end{proof}

We need one more lemma relating the random variables $\bm{\widetilde{Y}_t}$ and $\bm{Y'_t} \sim \text{Binomial}(t,\overline{q_t})$.

%\begin{sublemma}
\begin{lemma}
\label{lem:demoivre}
Let $\bm{Y_t(n)} := \text{Binomial}\left(\frac{n}{2}, \overline{q_t} + \delta\right) + \text{Binomial}\left(\frac{n}{2}, \overline{q_t} - \delta\right) + \text{Binomial}\left(t - n, \overline{q_t}\right)$ for any \textit{even} $n \in \{1,\ldots, t\}$.
Then, 
%as long as $T \geq T_2'$, 
there exists universal constant $\epsilon_c > 0$ such that for every $z \in [q^*t - \beta \sqrt{t}, q^*t + \beta \sqrt{t}]$, we have
\begin{align*}
\frac{\bbP(\bm{Y_t(n)} = z)}{\bbP(\bm{Y'_t} = z)} \geq \epsilon_c > 0.
\end{align*}
Recall that we defined $\bm{Y'_t} \sim \text{Binomial}(t,\overline{q_t})$.
\end{lemma}
% \end{sublemma}

Note that Lemma~\ref{lem:demoivre} immediately implies that 
\begin{align*}
\frac{\bbP(\bm{\widetilde{Y}_t} = z)}{\bbP(\bm{Y'_t} = z)} = \frac{\bbP(\bm{Y_t(n_t(z))} = z)}{\bbP(\bm{Y_t} = z)}\geq \epsilon_c > 0,  
\end{align*}
and we have thus related the original random variable $\bm{Y_t}$ to the Binomial random variable $\bm{Y'_t}$ through the constant $\epsilon_b := \epsilon_c$, which would complete the proof of Lemma~\ref{lem:changeofmeasure}.
Thus, it only remains to prove Lemma~\ref{lem:demoivre}, which we do below.

\proof{Proof of Lemma~\ref{lem:demoivre}.}
Our proof will critically use the classical de-Moivre-Laplace theorem, stated earlier.
To see how we can apply the de-Moivre-Laplace theorem to the denominator $\bbP(\bm{Y'_t} = z)$, we fix $q := \overline{q_t}$.
Then, note that since $z \in [tq^* - \beta \sqrt{t}, tq^* + \beta\sqrt{t}]$ and, from \eqref{eq:timeaverageoblivious},
we have $\overline q_t \in \{q^* - C/\sqrt{t},q^* + C/\sqrt{t}\}$.
Thus, we have 
$$z \in [t\overline q_t - (C-\beta)\sqrt{t}, t\overline q_t + (C + \beta)\sqrt{t}].$$
Designating $C_c := (\beta + C)$, we consider the choice of sequence $\{k_t = C_c \sqrt{t}\}_{t \geq 1}$.
This sequence clearly satisfies $k_t^3/t^2 \to 0$, and so we can directly apply the statement of the DeMoivre-Laplace theorem to get

\begin{align*}
 (1 - \epsilon_c) \cdot p\left(z;t \overline{q_t},t\overline{q_t}(1 - \overline{q_t})\right) \leq \bbP(\bm{Y'_t} = z) \leq (1 + \epsilon_c) \cdot p\left(z;t \overline{q_t},t\overline{q_t}(1 - \overline{q_t})\right),
\end{align*}
for all $t \geq t_{0,c}$ for all $qt - k_t \leq  z \leq qt + k_t$.
Further, we will adjust $t_{0,c}$ such that $t > t_{0,c} := t_{0,c}/q^*$.
% Henceforth, we will consider $T_2' := \max\{T_2'(c_1),T_2'(c_2)\}$.
% We will further assume that $T$ is large enough so that $T_2' < q^* T$.
Recall that $p(\cdot;\mu, \sigma^2)$ denotes the pdf of the normal distribution $\mathcal{N}(\mu, \sigma^2)$.

There are two cases to study depending on the value that $n$ takes.
The first one considers $n \leq t_{0,c}$. Noting that $t_{0,c}$ is a constant, in this case we can directly bound the ratio of pmfs.
First, we very crudely lower bound the numerator to get
\begin{align*}
\bbP(\bm{Y_t(n)} = z) &= \sum_{0 \leq k_1,k_2 \leq n/2,0 \leq k_3 \leq (t-n), k_1 + k_2 + k_3 = z} \binom{n/2}{k_1} \binom{n/2}{k_2} \binom{t-n}{k_3} \\
&(\overline{q_t} + \delta)^{k_1} \cdot (1 - \overline{q_t} - \delta)^{n/2 - k_1} \cdot (\overline{q_t} - \delta)^{k_2} \cdot (1 - \overline{q_t} + \delta)^{n/2 - k_2} \cdot (\overline{q_t})^{k_3} (1 - \overline{q_t})^{t - n - k_3} \\
&> \binom{t-n}{z - n} (\overline{q_t} + \delta)^{n/2} (\overline{q_t} - \delta)^{n/2} (\overline{q_t})^{z - n}(1 - \overline{q_t})^{t - z} ,
\end{align*}
where in the last inequality we considered only the point $k_1 = k_2 = n/2, k_3 = z - n$ in the sum.
(Note that this is a valid point as $z \leq t$ and $z - n \geq q^* t - t_{0,c} > 0$. The latter inequality follows because we assumed that $q^*T > t_{0,c}$.)
On the other hand, for the denominator we have
\begin{align*}
\bbP(\bm{Y'_t} = z) = \binom{t}{z} (\overline{q_t})^z (1 - \overline{q_t})^{t - z} ,
\end{align*}
and so we get, after some algebraic simplification,
\begin{align*}
\frac{\bbP(\bm{Y_t(n)} = z)}{\bbP(\bm{Y'_t} = z)} &> \frac{\binom{t - n}{z - n} \left(1 - \frac{\delta^2}{\overline{q_t}^2}\right)^{n/2}}{\binom{t}{z}} \\
&\geq \epsilon_c > 0 ,
\end{align*}
where the constant $\epsilon_c$ will depend on $\overline{q_t}, \delta,t_{0,c}$, but not on $t$.
Here, we have critically used $n \leq t_{0,c}$ to lower bound the term $\left(1 - \frac{\delta^2}{\overline{q_t}^2}\right)$ by such a constant, as well as noting that
\begin{align*}
\frac{\binom{t-n}{z - n}}{\binom{t}{z}} &= \frac{z}{t} \cdot \frac{z - 1}{t - 1} \ldots \frac{z - n + 1}{t - n + 1} \\
&\geq \left(\frac{ z - n + 1}{t - n + 1}\right)^{n} \\
&\geq (\epsilon_c)^n \geq (\epsilon_c)^{t_{0,c}} ,
\end{align*}
for some constant $\epsilon_c$ that is close to $q^*$.

Notice that the above crude argument does not work for the case where $n > t_{0,c}$, in particular, if it can grow indefinitely as a function of $t$, is less trivial. 
For this case, we make the following claim using the de-Moivre-Laplace theorem, under which it suffices to prove the lemma.
\begin{claim}\label{claim:demoivre}
There exists a constant $\epsilon_c \in (0,1)$ that can depend on $C_c$, but is independent of $(t, n)$, such that for $t_{0,c} \leq n \leq t$, 
%\footnote{\mycomment{Why cannot $n > T/2$? I think you meant to write $T_2' \leq n \leq T$.}}
we have
\begin{align}
\bbP(\bm{Y'_t} = z) &\leq (1 + \epsilon_c) \cdot p\left(z;t \overline{q_t},t\overline{q_t}(1 - \overline{q_t})\right) \label{eq:binomialdemoivre} \\
\bbP(\bm{Y_t(n)} = z) &\geq (1 - \epsilon_c) \cdot p\left(z; t \overline{q_t}, t \sigma^2(\overline{q_t}, n, t)\right) \label{eq:mixtureofbinomialsdemoivre}
\end{align}
 where 
 \[
     \sigma^2(\overline{q_t},n,t) := \frac{1}{t} \left(\frac{t}{2}(\overline{q_t} - \delta)(1 - \overline{q_t} + \delta) + \frac{n}{2}v^+(\overline{q_t};\delta) +  (t-n) \overline{q_t}(1 - \overline{q_t}))\right),
 \]
for any $z \in [tq^* - C_c\sqrt{t}, tq^* + C_c\sqrt{t}]$.
\end{claim}

First, notice that Claim~\ref{claim:demoivre} directly gives us our proof for the case where $n \geq t_{0,c}$.
To see this, consider the second case where $\overline{q_t} > 1/2$.
This gives us
\begin{align*}
\frac{\bbP(\bm{Y_t(n)} = z)}{\bbP(\bm{Y'_t} = z)} &\geq \frac{(1 - \epsilon_c)}{(1 + \epsilon_c)} \cdot \frac{p(z; t \overline{q_t}, t\sigma^2(\overline{q_t},n,t))}{p\left(z;t \overline{q_t},t\overline{q_t}(1 - \overline{q_t})\right)} \\
&= \frac{(1 - \epsilon_c)}{(1 + \epsilon_c)} \cdot \frac{\sqrt{2\pi \cdot t(\overline{q_t})(1 - \overline{q_t})}}{\sqrt{2\pi \cdot t\sigma^2(\overline{q_t},n,t)}} \cdot \frac{e^{-\frac{(z - t\overline{q_t})^2}{2 t \sigma^2(\overline{q_t},n,t)}}}{e^{-\frac{(z - t\overline{q_t})^2}{2 t\overline{q_t}(1 - \overline{q_t})}}} .
\end{align*}

First, we note that $\sigma^2(\overline{q_t}, n,t) \leq \frac{1}{4}$. Moreover, we know that $\overline{q_t} \in [q^* - \delta, q^* + \delta]$, and so we have
\begin{align*}
\frac{\sqrt{2\pi \cdot t\overline{q_t}(1 - \overline{q_t})}}{\sqrt{2 \pi \cdot t\sigma^2(\overline{q_t}, n,t)}} \geq \epsilon_c > 0 ,
\end{align*} 
where $\epsilon_c$ is a constant that depends only on $\delta$.
Thus, we get
\begin{align*}
\frac{\bbP(\bm{Y_t(n)} = z)}{\bbP(\bm{Y'_t} = z)} \geq \epsilon_c \cdot \frac{e^{-\frac{(z - t\overline{q_t})^2}{2 t\sigma^2(\overline{q_t},n,t)}}}{e^{-\frac{(z - t\overline{q_t})^2}{2 t\overline{q_t}(1 - \overline{q_t})}}} .
\end{align*}
Finally, we note that $z \in [tq^* - C_c\sqrt{t}, tq^* + C_c\sqrt{t}]$.
Thus, to lower bound the numerator we get
\begin{align*}
e^{-\frac{(z - t\overline{q_t})^2}{2 t\sigma^2(\overline{q_t},n,t)}} &\geq e^{-\frac{4C_c^2 \cdot t}{2 t\sigma^2(\overline{q_t},n,t)}} \\
&\geq e^{-\frac{4C_c^2}{2\sigma^2(\overline{q_t}, n,t)}} \geq \epsilon_c > 0 ,
\end{align*}
where we now use the fact that $\sigma^2(\overline{q_t}, n,t) \geq (\overline{q_t} + \delta)(1 - \overline{q_t} - \delta)$. 
Thus, we get $\overline{q_t} + \delta \leq q^* + 2\delta < 1$. 
Note that this constant $\epsilon_c$ will depend on $(q^*, \delta, C_c)$, but is independent of $t$.
For the denominator, we trivially have $e^{-\frac{(z - t\overline{q_t})^2}{2 t\overline{q_t}(1 - \overline{q_t})}} \leq 1$.
Putting all of these together, we get
\begin{align*}
\frac{\bbP(\bm{Y_t(n)} = z)}{\bbP(\bm{Y'_t} = z)} \geq \epsilon_c > 0 ,
\end{align*}
where $\epsilon_c$ is the product of all the above constants and thus depends on $(t_{0,c},q^*, C_c, \delta)$, but is independent of $t$.
Thus, given Claim~\ref{claim:demoivre}, we have proved Lemma~\ref{lem:demoivre}.
(A symmetric argument, which we omit, also works for the case $\overline{q_t} \leq 1/2$.)

It only remains to prove this claim, which we do below using the DeMoivre-Laplace theorem.
\proof{Proof of Claim~\ref{claim:demoivre}.}
% The inequalities in Equations~\eqref{eq:binomialdemoivre} and~\eqref{eq:mixtureofbinomialsdemoivre} essentially follow from the de-Moivre Laplace theorem, stated below:
As we noted above, Equation~\eqref{eq:binomialdemoivre} follows immediately from the statement of Theorem~\ref{thm:demoivrelaplace}.
% This is because we can substitute $q := \overline{q_T}$ and $Y_T := X \sim \text{Binomial}(T, \overline{q_T})$.
% Then, for some arbitrary constant $c_2 > 1$, from the definition of a limit we have
% \begin{align*}
% \bbP(Y_T = z) &\leq c_2 \cdot p\left(z;T \overline{q_T},T\overline{q_T}(1 - \overline{q_T})\right) \text{ for } T \geq T_2'(c_2) .
% \end{align*}
To prove Equation~\eqref{eq:mixtureofbinomialsdemoivre}, we need to do a little more work, but essentially we can exploit the mixture-of-binomials structure in the random variable $\bm{Y_t(n)} := \text{Binomial}\left(\frac{n}{2}, \overline{q_t} + \delta\right) + \text{Binomial}\left(\frac{n}{2}, \overline{q_t} - \delta\right) + \text{Binomial}\left(t - n, \overline{q_t}\right)$ for any \textit{even} $n \in \{1,\ldots, t\}$.

First, we consider the extreme case where the distribution is ``most different" from $\bm{Y_t}$, i.e. $n = t$.
In this case, note that $\bm{Y_t(t)} = \bm{Y_{t,1}} + \bm{Y_{t,2}}$ where $\bm{Y_{t,1}} \sim \text{Binomial}\left(\frac{t}{2}, \overline{q_t} + \delta\right)$ and $\bm{Y_{t,2}} \sim \text{Binomial}\left(\frac{t}{2}, \overline{q_t} - \delta\right)$, and the random variables $\bm{Y_{t,1}}$ and $\bm{Y_{t,2}}$ are independent.
Thus, we get
\begin{align*}
\bbP(\bm{Y_t(t)} = z) &= \sum_{y = z - t\overline{q_t} + C_c\sqrt{t}}^{t\overline{q_t} - C_c\sqrt{t}} \bbP(\bm{Y_{t,1}} = y) \bbP(\bm{Y_{t,2}} = (z-y)) \\
&\geq \sum_{y = \frac{t}{2} \cdot (\overline{q_t} + \delta) - C_c t^{5/9}}^{\frac{t}{2} \cdot (\overline{q_t} + \delta) + C_c t^{5/9}} \bbP(\bm{Y_{t,1}} = y) \bbP(\bm{Y_{t,2}} = (z-y)) .
\end{align*}

Now, observe that $y \in [t/2 \cdot (\overline{q_t} + \delta) - C_ct^{5/9}, t/2 \cdot (\overline{q_t} + \delta) + C_ct^{5/9}]$, and because we have assumed that $z \in [\overline{q_t} - C_c\sqrt{t}, \overline{q_t} + C_c\sqrt{t}]$, we also have $(z-y) \in [t/2 \cdot (\overline{q_t} - \delta) - C_c t^{5/9}, t/2 \cdot (\overline{q_t} - \delta) + C_ct^{5/9}]$ for slightly adjusted constant $C_c$.
Moreover, it is easy to verify that the sequence $\{k_t := C_c t^{5/9}\}_{t \geq 1}$ satisfies the conditions required for application of de-Moivre-Laplace theorem.
% \vidyacomment{The argument as presented below would work for any $t^p$ such that $1/2 < p < 2/3$.}

We denote $v^*(q;\delta) := (q + \delta)(1 - q - \delta)$ and $v^-(q;\delta) := (1 - q + \delta)(q - \delta)$ as the variances of the Bernoulli random variables with parameters $(q + \delta)$ and $(1 - q + \delta)$ respectively.
Therefore, for large enough $t \geq t_{0,c}$ (where $t_{0,c}$ will depend on $(\epsilon_c, \delta, q^*, C_c)$, and appropriately chosen constant $\epsilon_c \in (0,1)$, and the specified ranges of $(y, z)$, we get
\begin{align*}
\bbP(\bm{Y_{t,1}} = y) &\geq (1-\epsilon_c) \cdot p\left(y;\frac{t}{2}(\overline{q_t} + \delta), \frac{t}{2}v^+(\overline{q_t};\delta)\right) \\
\bbP(\bm{Y_{t,2}} = (z-y)) &\geq (1 - \epsilon_c) \cdot p\left((z-y);\frac{t}{2}(\overline{q_t} - \delta), \frac{t}{2}v^-(\overline{q_t};\delta))\right) ,
\end{align*}
and so we get
\begin{align*}
&\bbP(\bm{Y_t(t)} = z) \\
&\geq (1 - \epsilon_c)^2 \sum_{y = \frac{t}{2} \cdot (\overline{q_t} + \delta) - C_ct^{5/9}}^{\frac{t}{2} \cdot (\overline{q_t} + \delta) + C_ct^{5/9}} p\left(y;\frac{t}{2}(\overline{q_t} + \delta),\frac{t}{2}v^+(\overline{q_t};\delta)\right) \cdot p\left((z-y);\frac{t}{2}(\overline{q_t} - \delta), \frac{t}{2}v^-(\overline{q_t};\delta)\right) \\
&\stackrel{(\mathsf{i})}{\geq} (1 - \epsilon_c)^2 \cdot p\left(z;\frac{t}{2}(\overline{q_t} + \delta), \frac{t}{2}v^+(\overline{q_t};\delta)\right) \star p\left(z;\frac{t}{2}(\overline{q_t} - \delta), \frac{t}{2}v^-(\overline{q_t};\delta)\right) - 2(1 -\epsilon_c)^2 \cdot e^{-C_c t^{1/9}} \\
&= (1 - \epsilon_c)^2 \cdot p\left(z; t\overline{q_t}, \frac{t}{2}v^+(\overline{q_t};\delta) + \frac{t}{2}(\overline{q_t} - \delta), \frac{t}{2}v^-(\overline{q_t};\delta)\right) - 2(1 - \epsilon_c)^2 \cdot e^{-C_c t^{1/9}} \\
&\stackrel{(\mathsf{ii})}{\geq} \epsilon_c (1 - \epsilon_c) p\left(z; t\overline{q_t}, \frac{t}{2}v^+(\overline{q_t};\delta) + \frac{t}{2}(\overline{q_t} - \delta), \frac{t}{2}v^-(\overline{q_t};\delta)\right) .
\end{align*}

Here, inequality $(\mathsf{ii})$ follows for large enough $t \geq t_{0,c}$ noting that for the specified range of $z$, we have $p\left(z; t\overline{q_t}, \frac{t}{2}v^+(\overline{q_t};\delta) + \frac{t}{2}(\overline{q_t} - \delta), \frac{t}{2}v^-(\overline{q_t};\delta)\right) \geq \epsilon_c > 0$; and also that $e^{-C_c t^{1/9}}$ goes to $0$ as $t \to \infty$.
% \vidyacomment{This is a straightforward application of the definition of the limit, and I'm omitting the steps here. The reason I have said large enough $t$ again is because we are invoking the limit definition.}
Inequality $(\mathsf{i})$ follows by noting that
\begin{align*}
\sum_{y > \frac{t}{2} \cdot (\overline{q_t} + \delta) + C_ct^{5/9}} p\left(y;\frac{t}{2}(\overline{q_t} + \delta),\frac{t}{2}v^+(\overline{q_t};\delta)\right) &\leq \bbP\left(\bm{W} > C_c t^{1/18}\right) \\
&\leq e^{-C_c t^{1/9}} ,
\end{align*}
where $\bm{W}$ denotes the standard normal random variable, and we have overloaded notation in choices of $C_c$.
Similarly, we have
\begin{align*}
\sum_{y < \frac{t}{2} \cdot (\overline{q_t} - \delta) - C_ct^{5/9}}p\left(y;\frac{t}{2}(\overline{q_t} - \delta),\frac{t}{2}v^-(\overline{q_t};\delta)\right) \leq e^{-C_c t^{1/9}} .
\end{align*}
% \vidyacomment{Observe that I'm just using Gaussian tail bounds to bound the tail terms in the convolution sum.
% The high-level thing to keep in mind while reading this is that we allow deviations in $(y,(z-y))$ on the order of $t^{5/9}$ around their mean, but $z$ still deviates only on the order of $t^{1/2}$.
% }
%
From this, we get
\begin{align*}
&\sum_{y = \frac{t}{2} \cdot (\overline{q_t} + \delta) - C_ct^{5/9}}^{\frac{t}{2} \cdot (\overline{q_t} + \delta) + C_ct^{5/9}} p\left(y;\frac{t}{2}(\overline{q_t} + \delta),\frac{t}{2}v^+(\overline{q_t};\delta)\right) \cdot p\left((z-y);\frac{t}{2}(\overline{q_t} - \delta), \frac{t}{2}v^-(\overline{q_t};\delta)\right) \\
&\geq p\left(z;\frac{t}{2}(\overline{q_t} + \delta), \frac{t}{2}v^+(\overline{q_t};\delta)\right) \star p\left(z;\frac{t}{2}(\overline{q_t} - \delta), \frac{t}{2}v^-(\overline{q_t};\delta)\right) - 2e^{-C_c t^{1/9}} ,
\end{align*}
and plugging this in above gives inequality $(\mathsf{i})$.

Clearly, we have proved Equation~\eqref{eq:mixtureofbinomialsdemoivre} for this extreme case.
Let us now extend this extreme case more generally.
% From here-on, we designate $T_2' := \max\{T_2'(c_2), T_2'(c_1)\}$.
Recall that we assumed $n \geq t_{0,c}$.
In this case, by an identical argument to the above, we get
\begin{align*}
\bbP(\bm{Y_{n,1}} + \bm{Y_{n,2}} = z') \geq \epsilon_c(1 - \epsilon_c)^2  \cdot p\left(z; n\overline{q_t}, \frac{n}{2}v^+(\overline{q_t};\delta) + \frac{n}{2}(\overline{q_t} - \delta), \frac{n}{2}v^-(\overline{q_t};\delta)\right) .
\end{align*}

Thus, we can utilize a similar convolution argument as before to study $\bm{Y_t(n)} := \bm{Y_{n,1}} + \bm{Y_{n,2}} + \bm{Y_{(t-n),3}}$, where $\bm{Y_{(t-n),3}} \sim \text{Binomial}((t-n), \overline{q_t})$ and is independent from $\{\bm{Y_{n,1}},\bm{Y_{n,2}}\}$.
Thus, we get 
\begin{align*}
\bbP(\bm{Y_t(n)} = z) &= \bbP(\bm{Y_{n,1}} + \bm{Y_{n,2}} + \bm{Y_{(t-n),3}} = z) \\
&\geq \sum_{z' = n \overline{q_t} - C_cn^{5/9}}^{n \overline{q_t} + C_cn^{5/9}} \bbP(\bm{Y_{n,1}} + \bm{Y_{n,2}} = z') \bbP(\bm{Y_{(t-n),3}} = (z -z')) .
\end{align*}

% for any $z' \in [n\overline{q_t} - C_c\sqrt{n}, n\overline{q_t} + C_c\sqrt{n}]$.
There are two cases depending on the value of $(t-n)$:
\begin{enumerate}
\item $(t - n) \leq t_{0,c}$. 
In this case, because $\bm{Y_{(t-n),3}} \in \{0,\ldots, t_{0,c}\}$, we have
\begin{align*}
\bbP(\bm{Y_{(t-n),3}} = (z - z')) &\geq (\min\{\overline{q_t}, 1 - \overline{q_t}\})^{(t - n)} \\
&\geq (\min\{\overline{q_t}, 1 - \overline{q_t}\})^{t_{0,c}} .
\end{align*}

% where $c$ is a constant depending on $(\overline{q_T}, T_2')$.
Similarly, it is easy to verify that the normal pdf $p((z-z');(t-n)\overline{q_t}, (t-n) \overline{q_t}(1 - \overline{q_t}))$ is also bounded above by a constant $c'$ as $(z - z') \leq t_{0,c}$.
Thus, the ratio of the two is bounded by a universal constant $\epsilon_c$ \textit{ for all } $(z - z') \in \{0,\ldots,t_{0,c}\}$.
\item The second case is $(t-n) \geq t_{0,c}$.
Now, we note that $(z - z') \in [(t - n)\overline{q_t} - C_c\sqrt{n} - C_c\sqrt{t}, (t-n) \overline{q_t} + C_c\sqrt{n} + C_c\sqrt{t}]$, therefore, $(z - z') \in [(t-n) \overline{q_t} - C_c\sqrt{(t-n)}, (t-n) \overline{q_t} + C_c\sqrt{(t-n)}]$ for slightly adjusted constant $C_c$.
Then, since $(t - n) \geq t_{0,c}$ as well, we can apply the de-Moivre-Laplace theorem on the Binomial random variable $\bm{Y_{(t-n),3}}$ and show that
\begin{align*}
\frac{\bbP(\bm{Y_{(t-n),3}} = (z - z'))}{p((z-z');(t-n)\overline{q_t}, (t-n) \overline{q_t}(1 - \overline{q_t}))} \geq (1 - \epsilon_c)
\end{align*}
for the specified range on $(z - z')$.
\end{enumerate}

Thus, in both cases, for appropriately chosen $\epsilon_c > 0$ we get
\begin{align*}
\bbP(\bm{Y_t(n)} = z) &\geq \epsilon_c \sum_{z' = n \overline{q_t} - C_c n^{5/9}}^{n \overline{q_t} + C_cn^{5/9}} p\left(z';n \overline{q_t}, \frac{n}{2}v^+(\overline{q_t};\delta) + \frac{n}{2}v^-(\overline{q_t};\delta)\right) \cdot \\
&p\left((z-z'); (t-n) \overline{q_t}, (t-n) \overline{q_t} (1 - \overline{q_t}) \right) \\
&\geq \epsilon_c^2 \cdot p\left(z;n \overline{q_t}, \frac{n}{2}v^+(\overline{q_t};\delta) + \frac{n}{2}v^-(\overline{q_t};\delta)\right) \star \\
&p\left(z; (t-n) \overline{q_t}, (t-n) \overline{q_t} (1 - \overline{q_t}) \right) \\
&= \epsilon_c^2 \cdot p\left(z;t\overline{q_t}, \frac{n}{2}v^+(\overline{q
}_t;\delta) + \frac{n}{2}v^-(\overline{q_t};\delta) + (t-n) \overline{q_t}(1 - \overline{q_t})\right) ,
\end{align*}
% \mycomment{How do we get the second inequality above? I see that you are using the fact that the Binomial random variables are concentrated around mean. But can we say the same for the products above? This seems true but I am not able to see how.}
where the second inequality uses an identical argument as earlier again noting that $n \geq t_{0,c}$.
This completes the statement of the claim, and thus completes the proof.
% \vidyacomment{The application of the $\epsilon_c$'s here could be a bit confusing to read?}
%
% \Halmos
\endproof
% \end{proof}

Now that we have proved Claim~\ref{claim:demoivre}, we have completed the proof of Lemma~\ref{lem:demoivre}.
% \Halmos
\endproof
% \end{proof}

% \subsection*{Proof of Lemma~\ref{lem: ration_YtoZ}}
Finally, we prove Lemma~\ref{lem: ration_YtoZ} below.
\proof{Proof of Lemma~\ref{lem: ration_YtoZ}.}
   From the DeMoivre-Laplace theorem, there exists positive constant $\epsilon_d \in (0,1)$ and a sufficiently large $t_{0,d}$, such that for all $t > t_{0,d}$, we have
    \begin{align*}
\bbP(\bm{Y'_t} = z) &\geq (1 - \epsilon_d) \cdot p\left(z;t \overline{q_t},t\overline{q_t}(1 - \overline{q_t})\right),  \\
\bbP(\bm{Z'_t} = z) &\leq (1 + \epsilon_d) \cdot p\left(z;t q^*,t q^*(1 - q^*)\right).  
\end{align*}
Hence, we have
\begin{align*}
\frac{\bbP(\bm{Y'_t} = z)}{\bbP(\bm{Z'_t} = z)} &\geq \frac{(1 - \epsilon_d)}{(1 + \epsilon_d)} \cdot \frac{p(z; t \overline{q_t}, t\overline{q_t}(1 - \overline{q_t}))}{p\left(z;t{q^*},t{q^*}(1 - {q^*})\right)} \\
&= \frac{(1 - \epsilon_d)}{(1 + \epsilon_d)} \cdot \frac{\sqrt{2\pi \cdot t(\overline{q_t})(1 - \overline{q_t})}}{\sqrt{2\pi \cdot t{q^*_t}(1 - q^*)}} \cdot \frac{e^{-\frac{(z - t\overline{q_t})^2}{2 t \overline{q_t}(1 - \overline{q_t})}}}{e^{-\frac{(z - t{q^*})^2}{2 t{q^*}(1 - \overline{q^*})}}} .
\end{align*}
Now, we have
\[
    \frac{\sqrt{2\pi \cdot t(\overline{q_t})(1 - \overline{q_t})}}{\sqrt{2\pi \cdot t{q^*}(1 - q^*)}} \geq \max\left\{\frac{\sqrt{2\pi \cdot (q^* - \delta)(1 - q^* + \delta)}}{\sqrt{2\pi \cdot {q^*}(1 - q^*)}}, \frac{\sqrt{2\pi \cdot (q^* + \delta)(1 - q^* - \delta)}}{\sqrt{2\pi \cdot {q^*}(1 - q^*)}}\right\} > 0,
\]
for all $t$.
This is because $\overline{q_t} \in [q^* - \delta, q^* + \delta]$ and $\overline{q_t}(1 - \overline{q_t})$ being concave over this interval attains its minimum on the boundary. 
Further, we get
\[
    e^{-\frac{(z - t\overline{q_t})^2}{2 t \overline{q_t}(1 - \overline{q_t})}} \geq \max \left\{e^{-\frac{2C_d^2}{2 (q^* - \delta)(1 - q^* + \delta)}}, e^{-\frac{2C_d^2}{2 (q^* + \delta)(1 - q^* - \delta)}} \right\} > 0.
\]
Note that we have obtained bounds that do not depend on $t$.
Thus there exists a positive constant $\epsilon_d$ such that the statement in the lemma holds.
% \Halmos
\endproof
% \end{proof}

%\clearpage
% Bibliography:
\bibliographystyle{abbrvnat} % Citation style
\bibliography{Bib_Database}

\end{document}